\documentclass[aps,prd,reprint,showkeys,nofootinbib,superscriptaddress,notitlepage]{revtex4-1}
\usepackage{amsmath,graphicx,amssymb,multirow}

\usepackage[utf8]{inputenc}
\usepackage[unicode=true,
  bookmarks=false,   backref=false,    colorlinks=true,
  linktocpage=true,  citecolor=black,  linkcolor=black,
  urlcolor=black,    breaklinks=true
]{hyperref}

\newcommand{\GN}{G_{\rm N}}
\newcommand{\relG}{\Delta G/\GN}
\newcommand{\relGmax}{\relG |_\mathrm{max}}
\newcommand{\Rmax}{R_\mathrm{max}}

\newcommand{\be}{\begin{equation}}
\newcommand{\ee}{\end{equation}}

\begin{document}

\title{
The Hubble tension and fifth forces: a cosmic screenplay
}

\author{Marcus \surname{Högås}}
\email{marcus.hogas@fysik.su.se}
\affiliation{Oskar Klein Centre, Department of Physics, Stockholm University\\Albanova University Center\\ 106 91 Stockholm, Sweden}
\affiliation{Department of Mathematics, Stockholm University\\ 106 91 Stockholm, Sweden}

\author{Edvard \surname{Mörtsell}}
\email{edvard@fysik.su.se}
\affiliation{Oskar Klein Centre, Department of Physics, Stockholm University\\Albanova University Center\\ 106 91 Stockholm, Sweden}
 
\begin{abstract}
Fifth forces are ubiquitous in modified theories of gravity. In this paper, we analyze their effect on the  Cepheid-calibrated cosmic distance ladder, specifically with respect to the inferred value of the Hubble constant ($H_0$). We consider a variety of effective models where the strength, or amount of screening, of the fifth force is estimated using proxy fields related to the large-scale structure of the Universe. For all models considered, the local distance ladder and the \emph{Planck} value for $H_0$ 
agrees with a probability $\gtrsim 20 \, \%$, relieving the tension compared to the concordance model with data being excluded at $99 \, \%$ confidence. The alleviated discrepancy comes partially at the cost of an increased tension between distance estimates from Cepheids and the tip of the red-giant branch (TRGB). Demanding also that the consistency between Cepheid and TRGB distance estimates is not impaired, some fifth force models can still accommodate the data with a probability $\gtrsim 20 \, \%$. This provides incentive for more detailed investigations of fundamental theories on which the effective models are based, and their effect on the Hubble tension. 
\end{abstract}

\maketitle

\section{Introduction}
For almost a century, scientists have been engaged in the pursuit of precisely measuring the current expansion rate of the Universe, known as the Hubble constant ($H_0$). Its value has been a subject of controversy during most of this time. At present, there is a discrepancy between the SH0ES team value $H_0 = (73.0 \pm 1.0) \, \mathrm{km/s/Mpc}$ \cite{Riess:2021jrx} and the \emph{Planck} satellite data giving $H_0 = (67.8 \pm 0.5) \, \mathrm{km/s/Mpc}$ \cite{Planck2020}.
To infer $H_0$, the SH0ES team use a Cepheid-calibrated cosmic distance ladder while the \emph{Planck} value is based on the distance to the last scattering surface of the cosmic microwave background radiation. This discrepancy is commonly referred to as the Hubble tension.

The most immediate resolution is to attribute the tension to systematic errors. However, despite diligent efforts, this approach has not been entirely successful. Another possibility is that there is new physics beyond the cosmological standard model, see refs.~\cite{Abdalla:2022yfr,Schoneberg:2021qvd} for some examples. So far, there is no such consensus solution, although early dark energy appears to be one of the most popular models to date \cite{Karwal:2016vyq,Poulin:2018dzj,Poulin:2018cxd,Agrawal:2019lmo,Lin:2019qug,Kamionkowski:2022pkx}. Another proposal is that we are located in an underdense region of the Universe \cite{Keenan:2013mfa,Frith:2003tb,Whitbourn:2013mwa,Bohringer:2019tyj,Wong:2021fvu}. This results in an increased local expansion rate, explaining the ``high'' value of $H_0$ inferred by the SH0ES team \cite{Sundell:2015cza}. However, taking the full range of cosmological observations into account, this is ruled out \cite{Odderskov:2014hqa,Wu:2017fpr,Kenworthy:2019qwq,Camarena:2021mjr,Castello:2021uad}. 

In this paper, we explore an alternative approach based on fifth forces, first suggested in ref.~\cite{Desmond_2019}. A fifth force is the result of an extra degree of freedom and effectively leads to an increase in the gravitational force in certain environments, compared with the predictions of general relativity (GR). This can be modelled as an increase in the gravitational constant ($G$) compared with the Newtonian constant of gravitation ($\GN$). If $G > \GN$ in galaxies hosting both Type Ia supernovae (SNIa) and Cepheids but $G \simeq \GN$ in anchor galaxies with direct distance measurements, the SH0ES $H_0$ is biased to a high value. Thus, taking a fifth force into account can potentially harmonize the SH0ES value with \emph{Planck}. 

In ref.~\cite{Desmond_2019}, it was shown that an increase in $G$ relative to $\GN$ by $5\, \%$--$30\,\%$ in the host galaxies can alleviate the tension. In that work, the derived value of the Hubble constant was estimated by an effective rescaling. In the present paper, we infer the value of $H_0$ using a full statistical analysis.

We analyze three different models where the value of the fifth force is determined by the value of a phenomenological proxy related to the large-scale structure of the Universe. These are the externally sourced gravitational potential ($\Phi$), the externally sourced acceleration ($a$), and the externally sourced curvature ($K$), described in more detail below.\\

\noindent \textbf{Notation.} The numerical values of the Hubble constant ($H_0$) are given in units of $\mathrm{km/s/Mpc}$. Following ref.~\cite{Desmond_2019}, the externally sourced gravitational potential ($\Phi$) is given in units of $c^2$ where $c$ is the speed of light, the externally sourced acceleration ($a$) is given in units of $\mathrm{km/s^2}$, and the externally sourced curvature ($K$) is given in units of $1/\mathrm{cm}^2$. Concerning galaxy names, N4258 stands for NGC 4258 and U9391 stands for UGC 9391, etc. \newpage

\section{Executive summary}
In this paper, we follow the methods of the SH0ES team to set up the Cepheid-calibrated distance ladder with the addition that we take possible fifth force effects (parameterized by $\Phi$, $a$, and $K$) into account. Since the model parameter space allowed by the local distance ladder is infinite in these cases, it is not possible to obtain a constrained value for the Hubble constant, making the consistency between \emph{Planck} and the local distance ladder inconclusive at this level. 
By including a weight at each point in the parameter space representing the tension between the local distance ladder and the \emph{Planck} values for $H_0$, the allowed model parameter space becomes finite, allowing us to infer a constrained value for the Hubble constant from the local distance ladder, marginalized over the model parameters. For each model, we quantify the level of tension (or consistency) between the local distance ladder and \emph{Planck} by the $p$-value---the probability of the present data. 
\begin{itemize}
    \item The $\Phi$-model yields $H_0 = 68.0 \pm 1.3$ with a $p$-value of $p=0.44$.

    \item The $a$-model yields $H_0 = 69.4 \pm 1.3$ with a $p$-value of $p=0.25$.

    \item The $K$-model yields $H_0 = 69.9 \pm 1.3$ with a $p$-value of $p=0.18$. 
\end{itemize}
Compared with the standard model without a fifth force, which gives $p=0.01$, all fifth force models exhibit a fair consistency between \emph{Planck} and the local distance ladder. 

The degree of consistency between the local distance ladder and \emph{Planck} is generically largest for models where Cepheid-estimated galaxy distances are increased. At the same time, distances to the same galaxies estimated using the 
tip of the red-giant branch (TRGB)
decrease, since these are modified in the opposite direction in the presence of a fifth force. Thus, the eased tension comes partially at the cost of an impaired consistency between galaxy distance estimates based on Cepheids and the TRGB. 
However, since there is only a partial overlap between Cepheid and TRGB host galaxies, the degree of inconsistency depends on which galaxies are affected by the fifth force. Therefore, it is possible to obtain a $p$-value $\gtrsim 0.2$ while still satisfying the $95 \, \%$ confidence limit (CL) of the Cepheid versus TRGB distances.

Our results provide further support that a fifth force effective at galactic or large-scale structure scales can have beneficial properties with respect to the Hubble tension \cite{Desmond_2019}. Thus, it provides an incentive to study the foundational theories on which these effective fifth force models are based, with the aim of determining the degree to which these theories can alleviate the Hubble tension.

\section{Data description}
The present work requires the phenomenological proxy values obtained in ref.~\cite{Desmond_2019}. For this to cover all Cepheid anchor and host galaxies, we employ the same data sets and methodologies as previously described in refs.~\cite{mortsell2021hubble,Mortsell:2021tcx}, with necessary modifications to incorporate the effects of the fifth force.

To summarize, for the Large Magellanic Cloud (LMC) we utilize a distance modulus of $\mu_{\rm LMC}=18.477\pm 0.0263$ derived from double eclipsing binaries \citep{Paczynski:1996dj,Pietrzy_ski_2019,Riess_2019}. The distance to N4258 is determined through mega-maser observations to $\mu_{\rm N4258}=29.397\pm 0.032$  \citep{Reid_2019}. Data for the Milky Way (MW) Cepheids, including GAIA parallax measurements, are extracted from Table~1 in ref.~\cite{Riess_2021}.

For Cepheids in the LMC, we obtain the relevant data from Table~2 in ref.~\cite{Riess_2019}, while for Cepheids in M31 and beyond, we use the data in Table~4 in ref.~\cite{Riess:2016jrr}.

We obtain the SNIa peak magnitudes from Table~5 in ref.~\cite{Riess:2016jrr}, while TRGB data is sourced from ref.~\cite{Freedman_2019}.

The Hubble constant inferred from the local distance ladder is calculated using,
\be 
H_0=10^{M_{\rm B}/5+a_{\rm B}+5},
\label{eq:h0} 
\ee
with $a_{\rm B}=0.71273 \pm 0.00176$ being the intercept of the SNIa magnitude-redshift relation \citep{Riess:2016jrr} and $M_\mathrm{B}$ the SNIa \emph{B}-band peak absolute magnitude.

\begin{figure}[t]
    \centering
	\includegraphics[width=\linewidth]{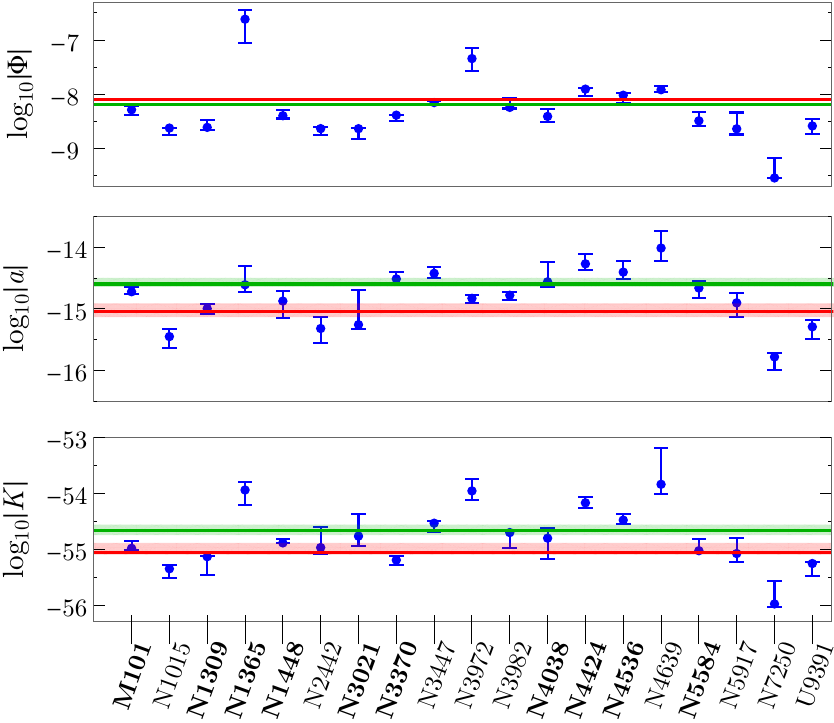}
	\caption{Values for the fifth force proxies for the anchor and SNIa host galaxies for some example models. The bold face galaxy names indicate the galaxies to which we have both Cepheid and TRGB distance estimates. Following the color coding of ref.~\cite{Desmond_2019}, the red line denotes the MW and the LMC (which exhibit the same proxy values) and the green line denotes N4258. The blue dots are the SNIa host galaxies. \emph{Top}: Values for the externally sourced gravitational potential ($\Phi$) for $\Rmax = 0.4 \, \mathrm{Mpc}$. \emph{Middle}: Values for the externally sourced acceleration ($a$) for $\Rmax = 18.1 \, \mathrm{Mpc}$. \emph{Bottom}: Values for the externally sourced curvature ($K$) for $\Rmax = 5.1 \, \mathrm{Mpc}$.
	\label{fig:screeningproxyvalues}}
\end{figure}

The fifth force proxies $(\Phi,a,K)$ are determined using the maps in ref.~\cite{Desmond_2017}, modelling the large scale structure out to distances of $200 \, \mathrm{Mpc}$. 
The proxy value for a specific galaxy achieves contributions from all sources within a certain cutoff radius ($\Rmax$) which designates the range of the fifth force. Some examples are shown in Fig.~\ref{fig:screeningproxyvalues}. Via ref.~\cite{Desmond_2017}, we have access to these values for five different cutoff radii, $\Rmax= [0.4 , 1.4 , 5.1 , 18.1 ,50] \, \mathrm{Mpc}$.
Proxy uncertainties are accounted for using a Monte Carlo method in the statistical data analysis. We simulate a sample of realizations where the values of the proxy parameters are drawn randomly from Gaussian distributions where the width is given by the error bars in Fig.~\ref{fig:screeningproxyvalues}. For a given proxy, each sample provides, at every point in the parameter grid: a best-fit $H_0$, the quality of the Cepheid distance ladder fit, the tension with the \emph{Planck} value for $H_0$, and the tension between Cepheid and TRGB distances.
The final value of $H_0$ and its corresponding uncertainty is given by a weighted average over the simulated samples and the parameter space, where the weight is given by the probability of each sample as derived from the corresponding Cepheid distance ladder $\chi^2$ and the tension with the value for $H_0$ derived from \emph{Planck} data.
We derive the $95\,\%$ confidence contour levels for the Cepheid distance ladder fit and the TRGB tension using the $5\,\%$ lowest percentile of the corresponding quantities across the random proxy samples. We refer to this method of inferring $H_0$ and confidence contours (in the parameter space) as the ``\emph{Planck}-weighted local distance ladder''.
The global minimum $\chi^2$ and corresponding $p$-values for each proxy model is given by the mean minimum $\chi^2$ across the simulations and its standard deviation.

When randomizing the proxy values, we investigate two extreme cases. One where the scatter is completely correlated between all galaxies, where for example a lower-than-average value for the proxy parameter is applied to all galaxies. In the other case, we assume that the uncertainty in the proxy values are completely uncorrelated between galaxies. The two alternatives give very similar results for the Hubble tension and TRGB consistency. As an example, for the $\Phi$-model with $\Rmax = 0.4 \, \mathrm{Mpc}$, the inferred Hubble constant value is $H_0 = 68.0 \pm 1.3$ both with correlated and uncorrelated errors in the proxy values and the $p$-values are $p=0.44$ and $p=0.41$. See Tab.~\ref{tab:CorrUncorr} for a complete comparative list. Here, as a default we assume that the errors in the proxy values are completely correlated. A discussion of the results in the case of uncorrelated proxy errors is contained in Appendix~\ref{sec:UncorrErr}.

\section{Methods I: (re-)calibrating the distance ladder}
The Cepheid-calibrated distance ladder consists of three steps, each possibly susceptible to the influence of a fifth force, potentially yielding a different value for $H_0$. To illustrate the effect, we express the Hubble constant as,
\begin{align}
    \label{eq:H0effects}
5\log_{10} H_0 = &5\log_{10} r(z)-5\log_{10} D^{\rm anch} \nonumber\\
&-\Delta m_{\rm Ceph}+\Delta m_{\rm SN},
\end{align}
where $r(z)$ can be approximated by $r(z) \simeq cz$ in the close Hubble flow and,
\begin{subequations}
    \begin{align}
\Delta m_{\rm Ceph}& = m_{\rm Ceph}^{\rm host}-m_{\rm Ceph}^{\rm anch},\\
\Delta m_{\rm SN}& = m_{\rm SN}^{\rm host}-m_{\rm SN}^{\rm flow},
\end{align}
\end{subequations}
see refs.~\cite{mortsell2021hubble,Mortsell:2021tcx}. Here, $\Delta m_{\rm Ceph}$ denotes a systematic offset in Cepheid magnitudes between SNIa host galaxies and anchor galaxies while $\Delta m_{\rm SN}$ represents a systematic offset in SNIa magnitudes between host galaxies and cosmic flow galaxies. In the absence of a fifth force, it is assumed that $\Delta m_{\rm Ceph} = \Delta m_{\rm SN} = 0$.

From eq.~\eqref{eq:H0effects}, we see that there are three ways that a fifth force can lower the value of $H_0$ (hence, easing the tension with the \emph{Planck} value), namely by:
\begin{enumerate}
\item Increasing the independent anchor distances, $D^{\rm anch}$.

\item Having $\Delta m_{\rm Ceph}>0$.
This condition applies when the fifth force is stronger in the SNIa host galaxies \cite{Desmond_2019}.
In this case, the host Cepheids appear brighter, necessitating a correction by raising $m_{\rm Ceph}^{\rm host}$.

\item Having $\Delta m_{\rm SN}<0$. If the SNIa in the Hubble flow experience a stronger fifth force than those in Cepheid hosts, they appear brighter, requiring a correction by raising $m_{\rm SN}^{\rm flow}$.
\end{enumerate}
A detailed description of the calibration process can be found in refs.~\cite{mortsell2021hubble,Mortsell:2021tcx}.

\subsection{Anchor distances}
In ref.~\cite{Desmond_2019}, the MW and N4258 were used as anchor galaxies. In our analysis, we extend the investigation by including a distance estimate to the LMC. In the following, we show that the distance anchors ($D^\mathrm{anch}$) are unaffected by a fifth force, that is, by a modified gravitational constant.\\ 

\noindent \textbf{MW.} The distances to the Cepheids in the MW are determined through observations of their parallax, providing a geometric measurement, independent of $G$.\\ 

\noindent \textbf{N4258.} The distance estimate to N4258 relies on observing the position, velocity (along the line of sight), and acceleration (along the line of sight) of water masers near its center. The model predictions for velocity and acceleration are based on the masers' Keplerian motion (plus relativistic corrections) where the gravitational constant always appear together with the mass of the central black hole ($M_\mathrm{BH}$), making only the combination $GM_\mathrm{BH}$ observationally constrained \cite{Humphreys:2013eja}. That is, any deviation from $\GN$ can be compensated by a corresponding change in $M_\mathrm{BH}$, so the fifth force does not influence the distance estimate to N4258.\\

\noindent \textbf{LMC.} The distance to the LMC is estimated using observations of detached eclipsing binaries (DEBs). The distance is inferred from the orbital velocity and photometric light curve of the system, which provide information about the physical size of the individual stars \cite{Paczynski:1996dj}. By considering the DEB temperatures, their luminosities can be determined, allowing for the distance estimation without assuming any value for $G$. Consequently, the estimated distance to the LMC remains unaffected by a modified gravitational constant.

\subsection{Cepheids}
Cepheid pulsation periods are correlated with their luminosities, making them standardizable candles. In the calibration process, one must correct for color and (potentially) metallicity. The pulsation period of a Cepheid is influenced by processes occurring in the star's envelope. In the presence of an unscreened envelope (i.e., $G > \GN$), the dynamics governing the pulsation is altered, as the free-fall time is reduced by a factor of $\sqrt{\GN/G}$ \citep{1980tsp..book.....C}. This suggests that the pulsation period is reduced by the same factor, which is also supported by more detailed models utilizing the linear adiabatic wave equation \citep{Sakstein:2013pda}. Consequently, in galaxies where Cepheid envelopes are unscreened, the period-luminosity relationship (PLR) is shifted compared to galaxies where the envelopes are screened (i.e., $G = \GN$). This shift in the PLR has the same effect as an increase in the Cepheid luminosity according to
\be
\label{eq:DeltaLA}
\Delta\log_{10} L =\frac{A}{2}\log_{10} \left(1+\frac{\Delta G}{G_{\rm N}}\right).
\ee
We adopt the value $A=1.3$, compliant with ref.~\cite{Desmond_2019}. The quantity $\relG$ denotes the relative increase in the gravitational constant, that is,
\begin{equation}
    \frac{\Delta G}{\GN} = \frac{G-\GN}{\GN}.
\end{equation}
The luminosity of a Cepheid is primarily determined by hydrogen burning in a thin shell surrounding the helium core. When this shell becomes unscreened, the star must consume more fuel to balance the enhanced gravitational force, resulting in higher luminosity. To account for this, a modified stellar structure code \citep{Sakstein:2019qgn} can be utilized to derive the relation
\be
\label{eq:DeltaLB}
\Delta\log_{10} L = B\log_{10} \left(1+\frac{\Delta G}{G_{\rm N}}\right),
\ee
where the coefficient $B$ depends on the mass of the Cepheid and whether it lies at the second or third crossing of the instability strip. We adopt $B = 4$ and assume that $\relG$ takes the same value in the envelope and the core.

To summarize, there are two fifth force effects that contribute to a shift in the PLR of Cepheids. The first arises from modified dynamics within the envelope, while the second stems from a modified burning rate in the vicinity of the core. The total shift in the PLR is the sum of the two effects, eq.~\eqref{eq:DeltaLA} and eq.~\eqref{eq:DeltaLB}.

\subsection{Type Ia supernovae}
\label{sec:DeltaMSN}
If a fifth force is present so that the white dwarf is unscreened, the effective gravitational force acting on it increases, resulting in a shift in the SNIa absolute magnitude \cite{Desmond_2019}. However, due to screening effects, generically $\relG \simeq 0$ in compact objects, with an increasing trend as the density decreases.
A typical mean density for a white dwarf is $10^6 \, \mathrm{g/cm^3}$ while it is $10^{-5} \, \mathrm{g/cm^3}$ for a typical Cepheid. Therefore, we set $\relG = 0$ for the SNIa, that is, no shift in their absolute magnitudes.

\subsection{TRGB consistency test}
\label{sec:ConsTest}
When the hydrogen at the core of a solar mass star is depleted, energy is primarily generated through hydrogen fusion in a shell surrounding the core. As the pressure and temperature of the core increase, for stars with masses less than $\simeq 1.8 \, M_\odot$, a rapid nuclear fusion process of helium, known as the helium flash, takes place. This results in a break in the luminosity evolution of the star, marking the tip of the red-giant branch. In the near-infrared $I$-band ($\sim 800 \, \mathrm{nm}$), the TRGB serves as a standard candle with an absolute magnitude of $M_I\simeq -4.0$. This can be employed as an alternative method for calibrating SNIa luminosities, ultimately leading to the determination of $H_0$ as discussed in ref.~\cite{Desmond:2020wep}. Here, we utilize the TRGB as a consistency test by comparing distances to galaxies inferred from both Cepheids and the TRGB.

\begin{figure*}[t]
    \centering
    \includegraphics[width=\linewidth]{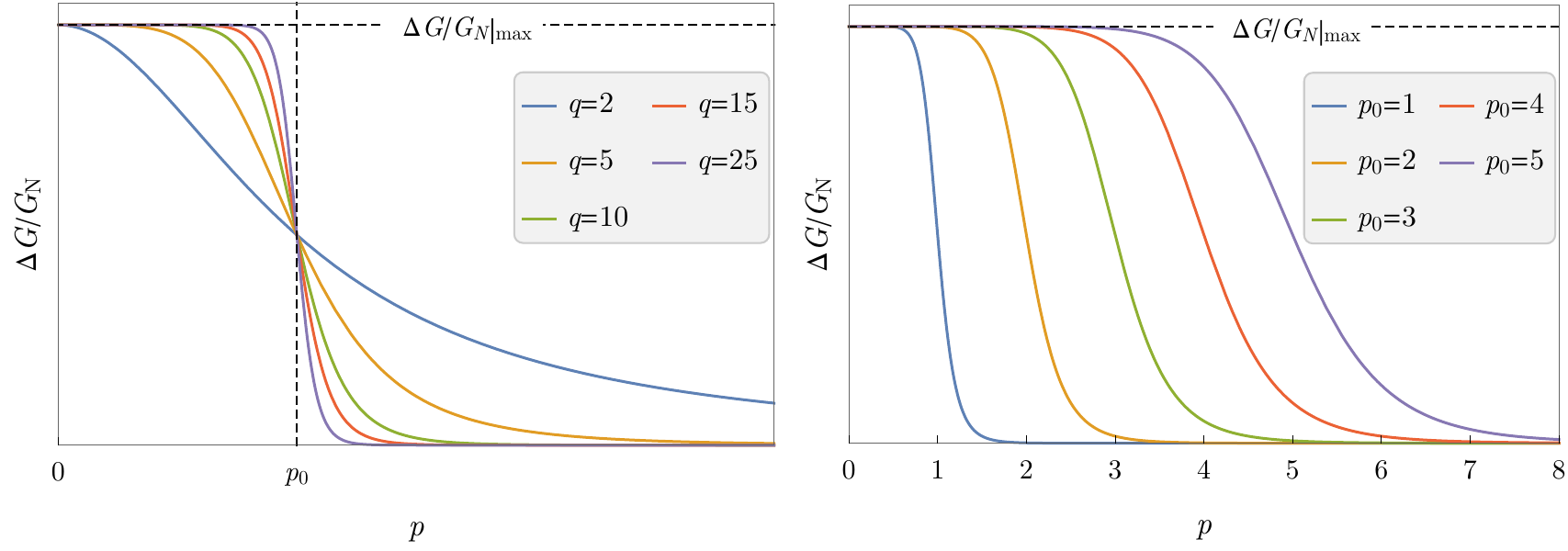}
    \caption{Examples of $\relG$ as a function of the proxy value $p$ as parameterized in eq.~\eqref{eq:delgvsp}. \emph{Left}: Varying $q$. Increasing $q$ yields a steeper transition. \emph{Right}: Varying $p_0$. Increasing $p_0$ pushes the transition to higher values of $p$.}
    \label{fig:relGmodel}
\end{figure*}

The luminosity of the red-giant branch stars (RGBs) is determined by a thin hydrogen shell surrounding the helium core. In the presence of a modified gravitational constant in this shell, the inferred distance is well approximated by the formula \citep{PhysRevD.101.129901},
\be
\label{eq:TRGBdistMod}
\frac{D_{\rm true}}{D_{\rm GR}}=1.021\sqrt{1-0.04663\left(1+\frac{\Delta G}{\GN}\right)^{8.389}}.
\ee
We assume that, for a given galaxy, $\relG$ takes the same value in the RGBs (eq.~\eqref{eq:TRGBdistMod}) as in the Cepheids (eqs.~\eqref{eq:DeltaLA},\eqref{eq:DeltaLB}).
Since distances derived from Cepheids and the TRGB are modified in opposite directions, significant fifth forces generically lead to systematic disagreements, making it possible to constrain the size of the fifth force.
Since the TRGB distance estimates in ref.~\cite{Freedman_2019} are calibrated using the RGBs in the LMC, we account for screening effects also in these.

\section{Methods II: Fifth force models}
In this section, we first describe how we calculate $G$ based on the value of the fifth force proxies and then describe the proxy models utilized in this paper.

\subsection{Mapping $G$}
To be compatible with solar-system tests of gravity, a fifth force must exhibit a screening mechanism suppressing its spatial variations on solar-system scales.\footnote{A constant rescaling of the gravitational force in the solar system can always be absorbed in a redefinition of $\GN$. The results in the present work remains unchanged with such a rescaling since we are only concerned with the relative variations in the fifth force.} In galactic environments however, the variations can be significant, leading to potentially observable effects. There is a plethora of screening mechanisms in the literature. For comprehensive reviews, see for example refs.~\cite{Khoury:2013tda,Joyce:2014kja,Brax:2021wcv}. Ideally, the value of the gravitational constant ($G$) should be calculated for each astrophysical object from the equations of motion of the theory. Such an approach is possible but computationally demanding (see e.g. ref.~\cite{Hogas:2023vim} for the case of symmetron screening). For practical purposes it can be useful to parameterize the strength of the fifth force by a proxy field ($p$) whose value is known observationally. In this paper, we adopt this approach and focus on proxy fields which are determined from the large-scale structure of the Universe, described in the section below.

The dependence of $G$ upon $p$ depends on the underlying fundamental gravity theory, although, to comply with current observational constraints, it should be constant in solar-system environments. In ref.~\cite{Desmond_2019}, the authors assume a stepwise mapping such that $G = \GN$ if $p$ is greater than some critical value ($p_\mathrm{crit}$) and $G = k \, \GN$ if $p < p_\mathrm{crit}$ where $k>1$ is some fixed constant. To account for a continuous dependence of $G$ on $p$, in this paper we set,
\begin{equation}\label{eq:delgvsp}
    \frac{\Delta G}{\GN} = \left. \frac{\Delta G}{\GN} \right|_\mathrm{max} \, \frac{1}{1 + (p/p_0)^q},
\end{equation}
where $\relGmax$ determines the maximal value of $\relG$, $q$ sets the width of the transition from $\relGmax$ to $0$, and $p_0$ determines the value of $p$ where the transition takes place. See Fig.~\ref{fig:relGmodel} for some examples.

If $q$ is small, the transition is slow, see Fig.~\ref{fig:relGmodel}. This results in all galaxies exhibiting similar values of $\relG$, resulting in little or no effect on the calibration of the distance ladder. Accordingly, we find that the greatest effect on $H_0$ is achieved when $q$ is large, that is, with a sharp transition between screened and unscreened galaxies. Thus, in the following, we set $q=500$ as a default value and show results for other values of $q$ in Appendix~\ref{sec:ComplRes}. 

In addition to the $(\relGmax,q,p_0)$ model parameters, $\Rmax$ denotes the range of the fifth force, that is, determining the radius out to which a source contributes to the fifth force.
 
\begin{figure*}[t]
    \centering
	\includegraphics[trim={0.1cm 0 0.7cm 0},width=0.34\linewidth,clip]{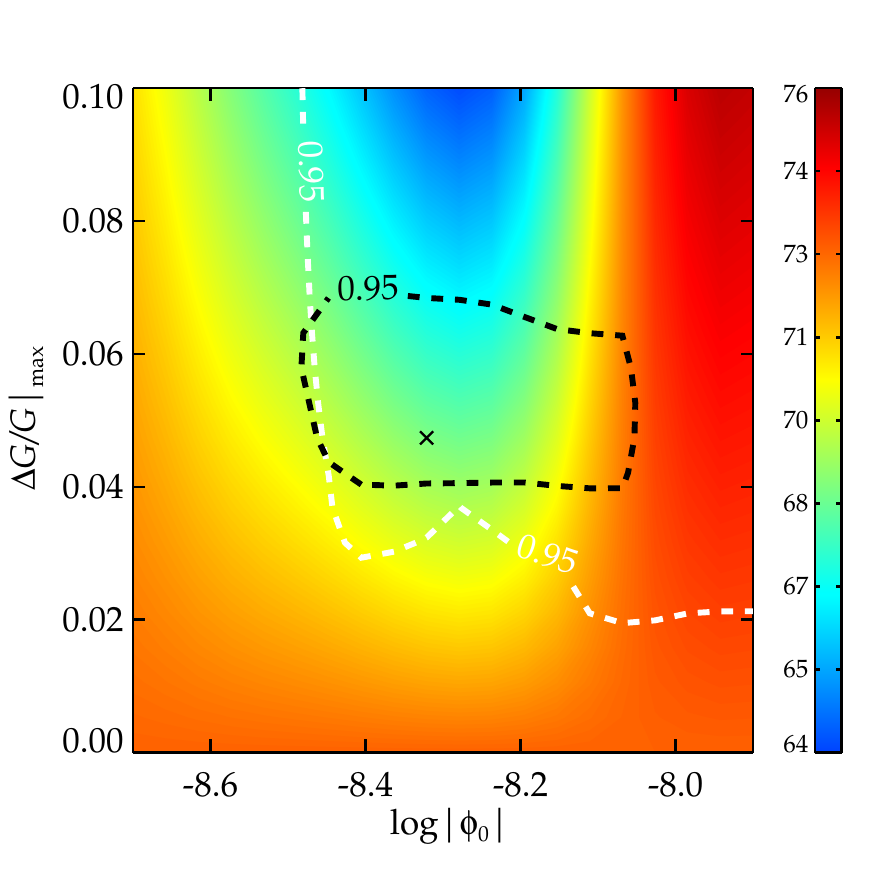}
    \includegraphics[trim={1.0cm 0 0.7cm 0},width=0.32\linewidth,clip]{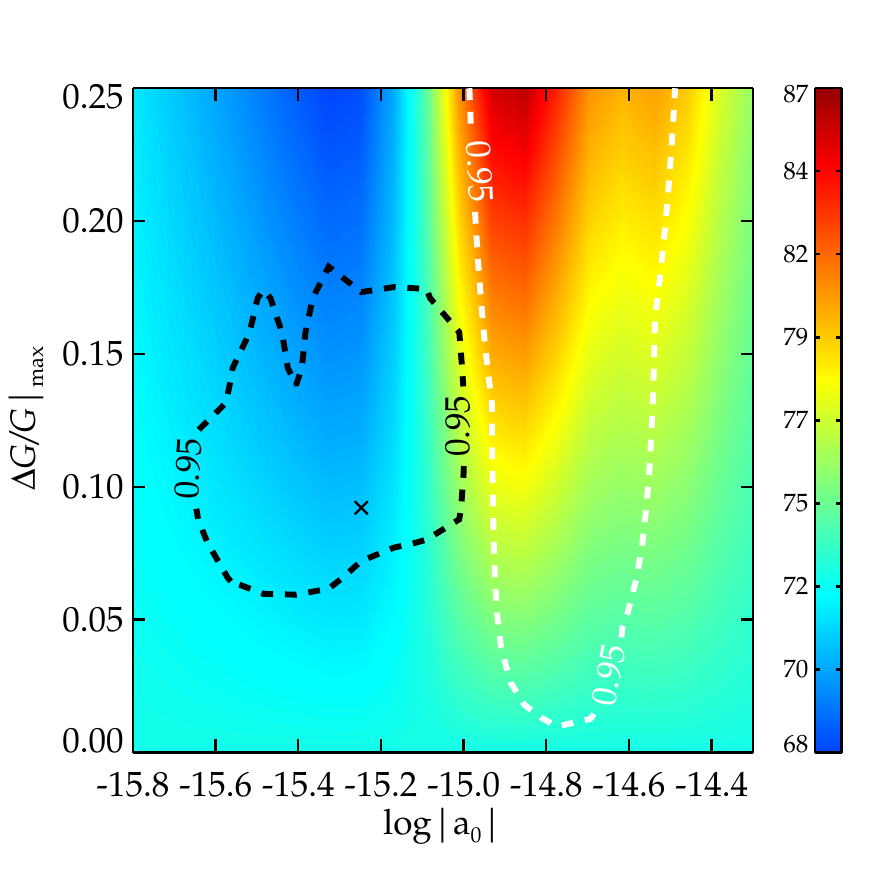}
    \includegraphics[trim={1cm 0 0.7cm 0},width=0.32\linewidth,clip]{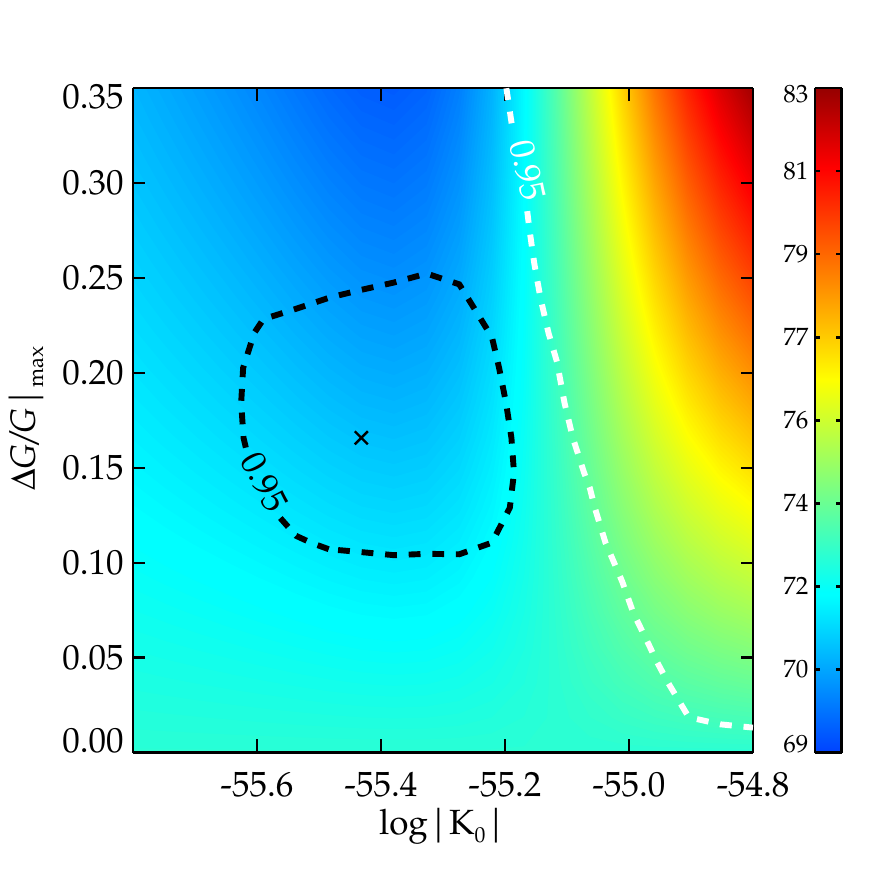}
    
	\caption{The color shade denotes the value of $H_0$ at each point in the parameter space. The black contours show the inferred $95 \, \%$ CLs from the \emph{Planck}-weighted local distance ladder. Everything above the white dashed curves is excluded by the Cepheid-TRGB consistency test with $95 \, \% $ confidence. The cross indicates the best-fit point, tabulated in Tab.~\ref{tab:resultsComp}. \emph{Left}: $\Phi$-screening with $\Rmax = 0.4 \, \mathrm{Mpc}$. \emph{Middle}: $a$-screening with $\Rmax = 18.1 \, \mathrm{Mpc}$. \emph{Right}: $K$-screening with $\Rmax = 5.1 \, \mathrm{Mpc}$.
	\label{fig:ScreeningWithPrior}}
\end{figure*}

\subsection{Screening proxies}
In this work, we analyze proxy models where the degree of screening is determined by the value of an observable related to the large-scale structure of the Universe. The proxy fields are the externally sourced gravitational potential ($\Phi$), the externally sourced acceleration ($a$), and the externally sourced curvature ($K$).

The values of the proxy fields are determined using the local screening maps in ref.~\cite{Desmond_2017}, from which we have access to the proxy values for five different values of the cutoff radius in the range $0.4 \, {\rm Mpc} \leq \Rmax \leq 50 \, {\rm Mpc}$.

In ref.~\cite{Cabre:2012tq}, it was shown that the externally sourced gravitational potential ($p=\Phi$) can be used as a proxy, parameterizing the degree of screening of an $f(R)$-model exhibiting chameleon screening. The degree of screening under a kinetic mechanism such as k-mouflage \cite{Babichev:2009ee} may be parameterized by the externally sourced acceleration ($p=a$) and Vainshtein screening models \cite{Dvali:2000hr,Dvali:2000xg,Nicolis:2008in,Babichev:2009jt,Deffayet:2009wt,Brax:2011sv} may be parameterized by the externally sourced curvature, quantified by the Kretschmann scalar ($p=K$).

\begin{figure*}[t]
    \centering
    \includegraphics[width=\linewidth]{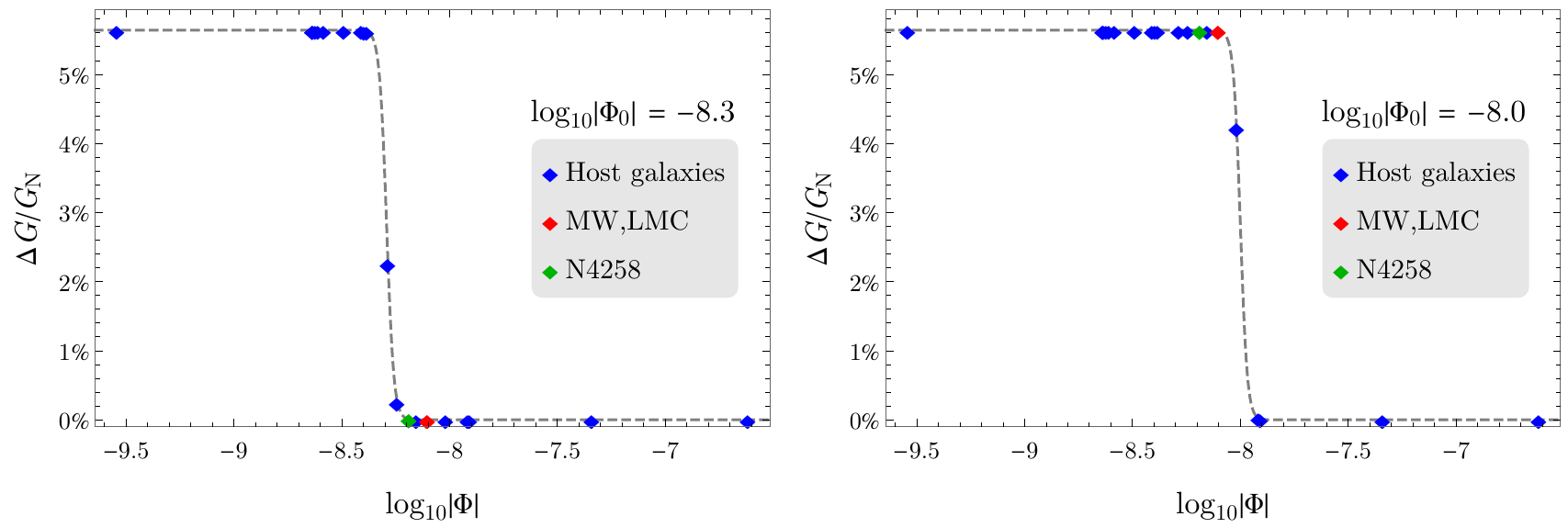}
    \caption{Examples of $\relG$ for host and anchor galaxies for a $\Phi$-screening model with $\Rmax=0.4 \, \mathrm{Mpc}$. \emph{Left}: $\log_{10} |\Phi_0|=-8.3$. Here, the anchor galaxies (MW, LMC, and N4258) are all screened while a significant fraction ($68 \, \%$) of the host galaxies are unscreened. In this case, the $H_0$ value is decreased. \emph{Right}: $\log_{10} |\Phi_0|=-8.0$. Here, the anchor galaxies are fully unscreened while some host galaxies are still screened. In this case, the inferred $H_0$ value increases.}
    \label{fig:transitionExample}
\end{figure*}

\section{Results}
In this section, we present the results of including the proxy fifth force models in the calibration of the cosmic distance ladder with the main goal to analyze how it affects the tension between the distance ladder and \emph{Planck}. In the accompanying figures, the black dashed lines are the $95 \, \%$ CLs obtained from the quality of the local distance ladder fit, and its agreement with the \emph{Planck} value of $H_0$. The white dashed lines show the $95 \, \%$ CLs obtained from comparing Cepheid and TRGB distances as described in Section~\ref{sec:ConsTest}.

For the screening models that we analyze here, each point in the parameter space yields a certain value of $H_0$. As seen in Fig.~\ref{fig:ScreeningWithPrior}, for small values of $\relGmax$, the Hubble constant approaches the value obtained without screening whereas increasing $\relGmax$ yields an increasing effect on $H_0$. When the proxy value $p_0$ is smaller than the minimum host galaxy value, all galaxies are screened and there is no effect on $H_0$. Increasing $p_0$ beyond the smallest galaxy value, some of the host galaxies become unscreened while the anchor galaxies remain screened, see Fig.~\ref{fig:transitionExample} (left panel) for an example. In this case, the inferred Hubble constant decreases compared to the standard case without a fifth force. Increasing $\relGmax$ increases the effect, making it possible to obtain arbitrarily low values of $H_0$. What prevents such an arbitrary decrease in the Hubble constant is the global fit of the distance ladder, which requires a full statistical analysis as implemented in this paper.
Increasing $p_0$ beyond the values of the anchor galaxies, the anchor galaxies are fully unscreened while some host galaxies are still screened, see Fig.~\ref{fig:transitionExample} (right panel). In this case, the Hubble constant increases compared with the standard case. Thus, we can understand the transition in Fig.~\ref{fig:ScreeningWithPrior} where the Hubble constant goes from being smaller than the standard SH0ES value to greater when increasing $p_0$. Increasing $p_0$ even further, beyond the greatest value among all galaxies, all galaxies become unscreened to the same degree and there is no effect on $H_0$. 

In general, the model parameter space allowed by the local distance ladder is infinite, making it impossible to obtain a constrained value for the Hubble constant by marginalizing over the model parameters, see Appendix~\ref{sec:NoH0Prior}. We therefore add the \emph{Planck} value ($67.8 \pm 0.5$) as a data point to be compared to the best fit local distance ladder value for $H_0$ when weighting parameter points during marginalization, allowing us to infer a constrained value for the Hubble constant from the local distance ladder by effectively making the allowed model parameter space finite. We refer to this method of obtaining $H_0$ as the ``\emph{Planck}-weighted local distance ladder''.

In the standard case, the Hubble tension manifests itself as a poor fit of the local distance ladder when compared to {\em Planck} data, that is, a low $p$-value ($p=0.01$)---the probability of the present data. For each fifth force proxy, we quantify the level of tension (or consistency) between the local distance ladder and \emph{Planck} by the $p$-value of the given model, where higher $p$-values indicate less tension. 

\begin{figure}[t]
    \centering
	\includegraphics[width=\linewidth]{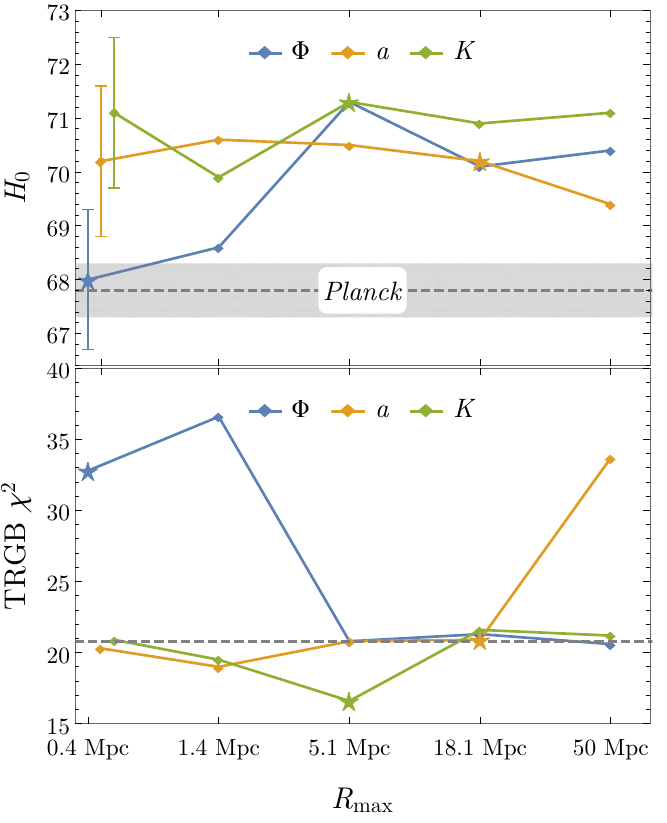}   
	\caption{\emph{Top}: Inferred value of $H_0$ from the \emph{Planck}-weighted local distance ladder. The sizes of the error bars are indicated by the leftmost points. The models featured in Fig.~\ref{fig:ScreeningWithPrior} are marked with stars. \emph{Bottom}: Consistency between Cepheid and TRGB distance estimates ($\chi^2$-value) for the best-fit screening models. The standard result (i.e., with no fifth force) is indicated by the dashed line. Some of the models improve the consistency between Cepheids and TRGB, while others (especially those with low values of $H_0$) worsen the Cepheid-TRGB consistency.
	\label{fig:ResultsAllModels}}
\end{figure}

In Fig.~\ref{fig:ResultsAllModels}, we present the values of $H_0$ inferred from the \emph{Planck}-weighted local distance ladder and in Tab.~\ref{tab:resultsH0prior}, we present numerical results for selected models and quantify the performance of the screening models compared with the standard case. A comprehensive list of results is given in Appendix~\ref{sec:ComplRes}.

The model which exhibits the lowest local value of $H_0$ (inferred from the \emph{Planck}-weighted local distance ladder) is $\Phi$-screening with $\Rmax = 0.4 \, \mathrm{Mpc}$. In this case, $H_0 = 68.0 \pm 1.3$, with a $p$-value of $p=0.44$. 
Hence, for this model the local distance ladder is consistent with \emph{Planck}.
Note that the consistency between the local $H_0$ value and the \emph{Planck} value is mainly due to the best-fit value of the Hubble constant changing, rather than an increased uncertainty in $H_0$.
For externally sourced acceleration ($a$), $\Rmax = 50 \, \mathrm{Mpc}$ provides the best consistency (lowest tension) between the distance ladder and \emph{Planck} with $p=0.25$ and $H_0 = 69.4 \pm 1.3$, see Tab.~\ref{tab:resultsComp}.
For externally sourced curvature ($K$), $\Rmax = 1.4 \, \mathrm{Mpc}$ provides the lowest tension, with $p=0.18$ and $H_0 = 69.9 \pm 1.3$, see Tab.~\ref{tab:resultsComp}. 

\begin{table}
  \centering
  \begin{tabular}{c|c|c|c|c}
    \hline \hline
    & GR & $\Phi (0.4 \, \mathrm{Mpc})$ & $a (18.1 \, \mathrm{Mpc})$ & $K (5.1 \, \mathrm{Mpc})$\\
    \hline
    
    $H_0$ & $73.2\pm 1.3^*$ & $68.0\pm 1.3$ & $70.2\pm 1.3$ & $71.3\pm 1.3$\\
    
    $p$-value & $0.01$ & $0.44$ & $0.15$ & $0.07$\\

    $\chi^2$ & $1747$ & $1622$ & $1674$ & $1698$\\
    
    TRGB $\chi^2$ & $21$ & $33$ & $21$ & $17$\\
\hline \hline
  \end{tabular}
  \caption{Effect of some example screening models on the Hubble tension. The tabulated $H_0$ value is inferred from the \emph{Planck}-weighted local distance ladder. $^*$The exception is the standard distance ladder value for $H_0$ for which no marginalization is required.}
  \label{tab:resultsH0prior}
\end{table}

Unlike the dependence on $q$, there is no simple way to predict the dependence of the inferred $H_0$ on $\Rmax$ in Fig.~\ref{fig:ResultsAllModels}. The irregularities is due to the fact that the proxy field values depend on the distribution of sources within the cutoff radius. Thus, how the derived $H_0$ changes with $\Rmax$ depends on how the distribution of sources changes in the anchors galaxies compared with the host galaxies. For example, if the number of sources increase more in the anchor galaxies than in the host galaxies as $\Rmax$ changes, the anchor galaxies become be more screened compared with the host galaxies and the best-fit $H_0$ value decreases. 

We conclude that fifth forces can alleviate the Hubble tension. However, this comes partially at the cost of worsening the consistency between Cepheid and TRGB distance estimates.
With no fifth force, the comparison of Cepheid and TRGB distances gives $\chi^2 \simeq 21$. Generically, the models with the greatest $p$-values (and lowest $H_0$) exhibit the greatest degree of tension between Cepheid and TRGB distances; compare the top and bottom panels of Fig.~\ref{fig:ResultsAllModels}. For example, the $\Phi$-model with the greatest $p$-value ($p=0.44$) yields $\chi^2 \simeq 33$ for the TRGB consistency test. The $a$-model with $\Rmax = 18.1 \, \mathrm{Mpc}$, on the other hand exhibits a slightly lesser (but still significant) degree of easing of the Hubble tension with $p=0.15$. In this case, the TRGB $\chi^2$ stays at $\chi^2 \simeq 21$.
Thus, the eased Hubble tension partially (but not completely) comes at the cost of an impaired consistency with TRGB. Note however that there is no simple one-to-one correspondence between reduced Hubble tension and worsened consistency with TRGB. This can be illustrated by the fact that two screening models with similar effects on the Hubble tension can have vastly different effects on the consistency with TRGB, cf. Tab.~\ref{tab:resultsComp}.
The reason is that there is only a subset of galaxies to which we have simultaneous Cepheid and TRGB distance estimates. So, if the recalibration of $H_0$ is largely caused by fifth force effects in galaxies to which we do not have TRGB distance estimates, the Hubble tension can be reduced while leaving the TRGB consistency unaffected. This is of course an idealized (arguably contrived) case and we discuss this effect further in Section~\ref{sec:Discussion}. At this point, we simply note that there are screening models that both stay within the $95 \, \%$ TRGB confidence limit at the same time as $p \gtrsim 0.2$, that is, easing the tension to a significant degree.

\section{Discussion}
\label{sec:Discussion}
\subsection{Consistency with TRGB}
We have shown that large-scale structure screening effects on the local distance ladder can potentially alleviate the Hubble tension. However, this comes partially at the cost of worsening the consistency between Cepheid and TRGB distance estimates. This can be most easily understood in the case where the anchor galaxies are screened. If some of the host galaxies to which we have both Cepheid and TRGB distance estimates are unscreened, then the Cepheid distances to these galaxies are underestimated while the distances inferred from the TRGB are overestimated. In this case, an increased fifth force increases the inconsistency between the Cepheid and TRGB distance estimates at the same time as it decreases the value of $H_0$, making it more compatible with \emph{Planck}.

As an example, we take the best-fit $\Phi$-screening model with $\Rmax = 0.4 \, \mathrm{Mpc}$ (i.e., $\relGmax \simeq 0.05$, $\log_{10} |\Phi_0| \simeq -8.3$). In this case, the MW and the LMC have $\relG \sim 10^{-7}$ while for N4258 $\relG \sim 10^{-4}$, so all anchor galaxies are screened. At the same time, there are seven unscreened ($\relG > 0.01$) host galaxies to which we have both Cepheid and TRGB distance estimates. Since the consistency between these distances is impaired by the unscreening of these host galaxies, the $\chi^2$-value for the TRGB consistency test increases.

On the other hand, if all Cepheid+TRGB host galaxies are screened, then the consistency is unaffected. Note that this can occur even if some of the host galaxies (to which we do not have TRGB distance estimates) are unscreened, thus yielding a lower $H_0$ value while passing the TRGB consistency test. This is the case for some of the best-fit screening models, including for example the $a$-screening model with $\Rmax = 5.1 \, \mathrm{Mpc}$, cf. Fig.~\ref{fig:ResultsAllModels}.
In this case, all Cepheid+TRGB host galaxies are screened ($\relG < 0.01$), including LMC, which is the TRGB anchor galaxy. Hence, the TRGB consistency is unchanged. At the same time, some of the host galaxies to which there are no TRGB distance estimates, exhibit a significant amount of unscreening, for example $\relG = 0.13$ for N7250. Ultimately, this results in a decrease in $H_0$ to $70.5 \pm 1.3$.
The only thing that prevents such a scenario from solving the Hubble tension by increasing $\relGmax$ further is the quality of the global distance ladder fit for such a model.\footnote{One may also think that such as solution, relying on the unscreened fifth force in one or a few specific host galaxies, such as N7250, would be too contrived or too sensitive to complementary observational constraints to be interesting.}

In fact, the quality of the fit of TRGB versus Cepheid distances can even be improved with the proxy screening models. To understand how, we begin by reminding that if the anchor galaxies are unusually screened compared to the host galaxies, the Cepheid-calibrated distances are underestimated and the TRGB-calibrated distances are overestimated, and vice versa if the anchor galaxies are unusually unscreened. Here, the relevant host galaxies are those common to both the Cepheid and TRGB distance ladders (marked in bold face in Fig.~\ref{fig:screeningproxyvalues}). Importantly, the Cepheid anchor galaxies are MW, LMC, and N4258 while the TRGB anchor galaxy is LMC, so the anchor galaxies of the two distance ladders are only partially overlapping. This means that the Cepheid anchor galaxies can be (on average) unusually screened while the TRGB anchor galaxy (LMC) can be unusually unscreened, compared with the host galaxies. Of course, this can only happen to a certain degree since the LMC is shared as an anchor between both distance ladders. 

\begin{figure}[t]
    \centering
	\includegraphics[width=\linewidth]{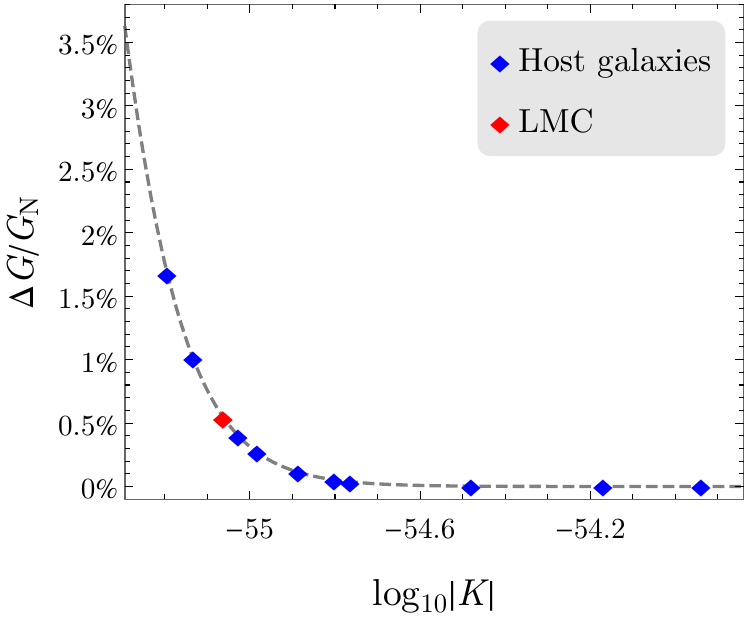}
	\caption{Modified gravitational constant for the LMC and the host galaxies to which we obtain both Cepheid and TRGB distance estimates. This is for a $K$-screening model with $\Rmax = 5.1 \, \mathrm{Mpc},\relG = 0.17,\log_{10} |K_0| = -55.4$. Note that the LMC is more unscreened than the majority of the host galaxies.
	\label{fig:trgbexample}}
\end{figure}

As an example we study the best-fit $K$-screening model with $\Rmax = 5.1 \, \mathrm{Mpc}$. The proxy values for the galaxies of this model are displayed in Fig.~\ref{fig:screeningproxyvalues} (bottom). From this figure, we see that a transition at $\log_{10} |K_0| = -55.4$ makes most of the galaxies screened while a couple of host galaxies (N1309 and N3370) are unscreened. For N4258 we have $\relG \sim 10^{-4}$ while the MW and the LMC are on the verge of being unscreened, with $\relG \simeq 0.005$. This causes the Cepheid anchor galaxies to be unusually screened compared with the host galaxies. On the other hand, the LMC is more unscreened than the average host galaxy, as can be seen in Fig.~\ref{fig:trgbexample}. 
Altogether, this increases the distance estimates to the host galaxies both for the Cepheid-calibrated distance ladder and the TRGB. The quality of the fit of TRGB versus Cepheid distances is improved from $\chi^2 \simeq 21$ without screening to $\chi^2 \simeq 17$ with this screening model. 

\subsection{Comparative study}
As mentioned previously, the work presented in this paper is an extension of that in ref.~\cite{Desmond_2019}, including:
\begin{itemize}
    \item A full statistical data analysis.

    \item Adopting the LMC as an anchor galaxy.

    \item Not assuming, \emph{a priori}, that the anchor galaxies are screened, but letting the model decide. Thereby letting $p_0$ be a free model parameter.\footnote{This is the case for the Cepheid-calibrated distance ladder as well as the TRGB consistency test.}

    \item A continuous dependence of $\relG$ on the proxy value.
\end{itemize}
Each of these differences has the potential of altering the results of ref.~\cite{Desmond_2019}. Nevertheless, we share the main conclusion, that the tension with \emph{Planck} can be eased below $2 \sigma$ while still being compatible with the TRGB consistency test. 

A quantitative comparison reveals that we generically obtain a greater tension between \emph{Planck} and the Cepheid-calibrated distance ladder in the current work with an average of $\simeq +1.7\sigma$ greater tension in our analysis compared with ref.~\cite{Desmond_2019}.
This is based on a comparison between the results in Table IV in ref.~\cite{Desmond_2019} with our results presented in Appendix~\ref{sec:NoH0Prior} where we, to comply with ref.~\cite{Desmond_2019}, analyze the lowest value of $H_0$ allowed by the distance ladder alone (i.e., when the Planck $H_0$ is not used to weight the points in the parameter space).

As an example, for the $K$-screening model with $\Rmax = 5.1 \, \mathrm{Mpc}$, the smallest $H_0$ allowed by TRGB distances is in $3.4 \sigma$ tension with \emph{Planck} while the corresponding value in ref.~\cite{Desmond_2019} is $1.5 \sigma$. The difference can be understood by recalling that we let $\relGmax$ and $p_0$ be free parameters. In this case, there are points in the parameter space where the TRGB fit is significantly improved compared with a standard model without a fifth force. For example, the $\chi^2$-value for the TRGB consistency can be as low as $\chi^2_\mathrm{min} \simeq 16.7$ for this screening model while $\chi^2 \simeq 20.8$ is the value without a screening model. In ref.~\cite{Desmond_2019} it is assumed that all anchors galaxies are screened, hence assuming $\chi^2_\mathrm{min} \simeq 20.8$ instead of $\chi^2_\mathrm{min} \simeq 16.7$. This means that the requirement to stay within the TRGB $95 \, \%$ CL, is more restrictive in our analysis, thus explaining the reduced easing of the Hubble tension in our analysis compared with ref.~\cite{Desmond_2019}.

\subsection{Theoretical foundations}
In the present work, we have studied effective fifth force models where the degree of screening is parameterized by a phenomenological proxy value related to the large-scale structure of the Universe. We have shown that some of these models can ease the Hubble tension to probabilities $\gtrsim 20\%$ while staying within the $95 \, \%$ CL for the TRGB consistency test. This motivates further study of these screening models, in particular their theoretical motivation. In the literature, it has been argued that many of the common screening mechanisms can be parameterized by these proxy values, see for example refs.~\cite{Khoury:2013tda,Desmond_2019,Brax:2021wcv}. 

Typically, screening parameterized by the externally sourced gravitational potential ($\Phi$) is associated with thin-shell mechanisms such as the chameleon, symmetron, and dilaton.\footnote{For the sake of completeness, it should be noted that the symmetron model only exhibits a thin-shell mechanism under certain circumstances, depending on the environment and the theory parameters. See ref.~\cite{Hogas:2023vim} for details.} In ref.~\cite{Cabre:2012tq}, it was shown that the chameleon screening induced by $f(R)$ gravity can be represented by the proxy $\Phi$. However, for an observationally viable chameleon mechanism, all the distance ladder galaxies are screened and accordingly there is no effect on the Hubble tension \cite{Desmond_2019,Jain:2012tn}.  Concerning the symmetron model, it does not have beneficial properties with respect to the Hubble tension, as shown in ref.~\cite{Hogas:2023vim}. More generally, it has been argued that the observational constraints from other gravity probes prohibit thin-shell mechanisms from affecting the cosmic distance ladder calibration of $H_0$ \cite{Desmond_2019}.

The degree of screening under a kinetic mechanism such as k-mouflage \cite{Babichev:2009ee} may be parameterized by the externally sourced acceleration ($p=a$) and Vainshtein screening models \cite{Dvali:2000hr,Dvali:2000xg,Nicolis:2008in,Babichev:2009jt,Deffayet:2009wt,Brax:2011sv} may be parameterized by the externally sourced curvature, quantified by the Kretschmann scalar ($p=K$).
Complementary observational constraints on kinetic and Vainshtein screening mechanisms seem to prohibit the influence of a fifth forces on the calibration of the cosmic distance ladder also for these models (see e.g. refs.~\cite{deRham:2016nuf,Sakstein:2017bws,Sakstein:2017xjx,Desmond_2019}). However, as in the case of symmetron screening \cite{Hogas:2023vim}, each theory needs close individual examination to establish whether it provides an observationally viable screening model that have beneficial properties with respect to the Hubble tension. 
It should also be stressed that a recalibration of the cosmic distance ladder does not necessarily involve a fifth force but can in principle be induced by any other phenomenon that has the effect of introducing systematic differences in the Cepheid period-luminosity relation between different galaxies.

\begin{acknowledgements}
EM acknowledges support from the Swedish Research Council under Dnr VR 2020-03384. Thanks to Harry Desmond for sharing data files from ref.~\cite{Desmond_2019}.
\end{acknowledgements}

\appendix

\begin{figure*}[t]
    \centering
	\includegraphics[trim={0.1cm 0 0.7cm 0},width=0.34\linewidth,clip]{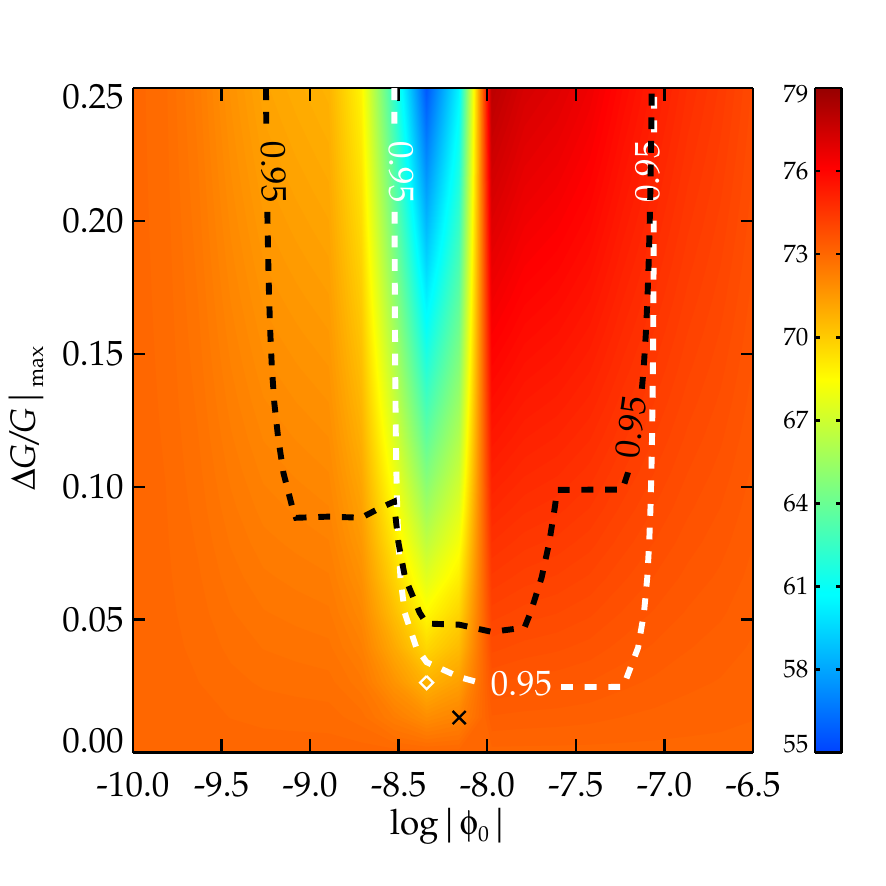}
    \includegraphics[trim={1.0cm 0 0.7cm 0},width=0.32\linewidth,clip]{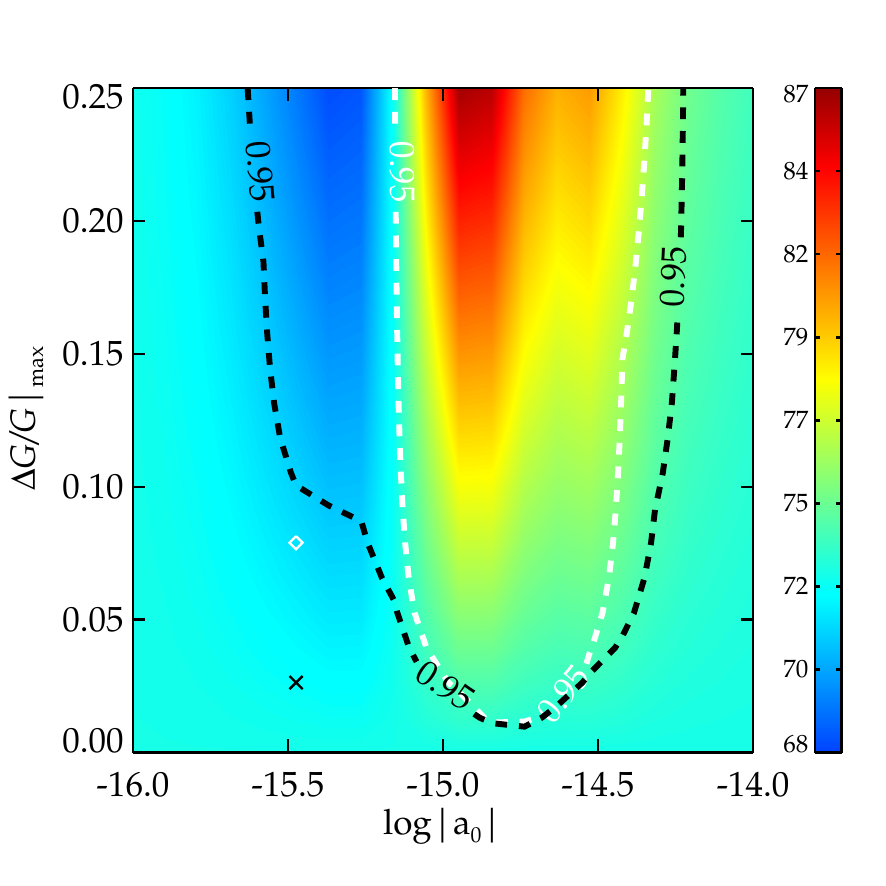}
    \includegraphics[trim={1cm 0 0.7cm 0},width=0.32\linewidth,clip]{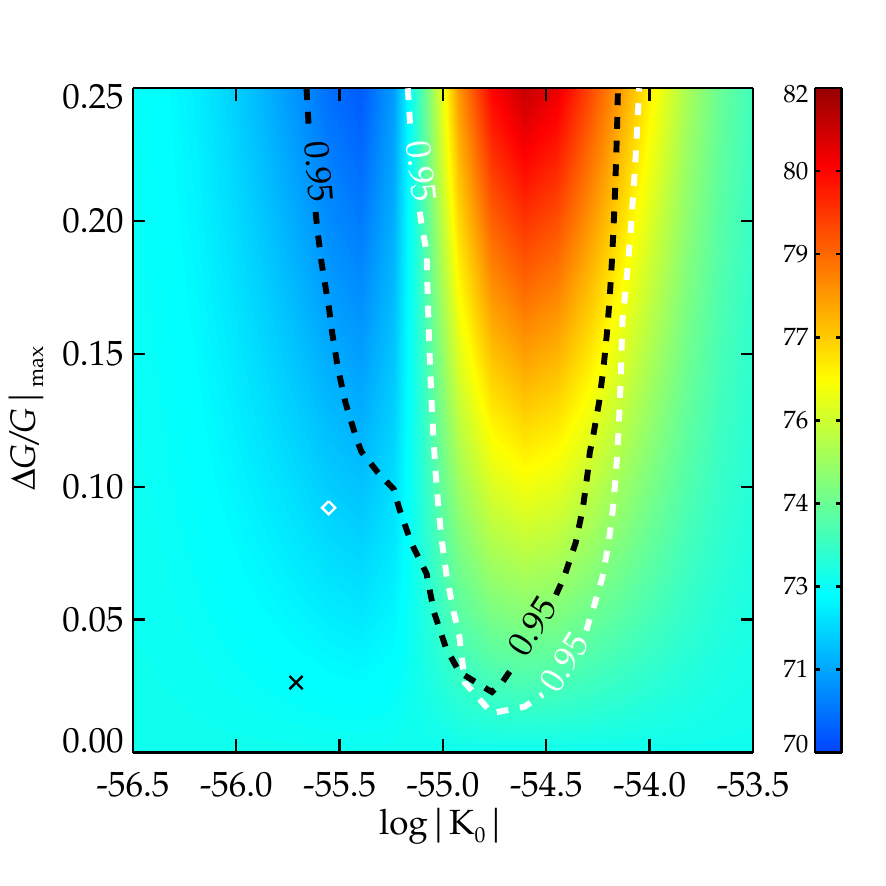}
    
	\caption{Inferred $H_0$ as a function of the model parameters. Black contours: $95 \, \%$ CLs from the Cepheid-calibrated distance ladder (no weighting with the \emph{Planck} $H_0$). White contours: $95 \ \%$ CLs obtained from comparing Cepheid and TRGB distances. The cross indicates the best-fit point and the diamond the point with the lowest $H_0$ value compatible with the $95 \, \%$ CLs. \emph{Left}: $\Phi$-screening with $\Rmax = 0.4 \, \mathrm{Mpc}$. \emph{Middle}: $a$-screening with $\Rmax = 18.1 \, \mathrm{Mpc}$. \emph{Right}: $K$-screening with $\Rmax = 5.1 \, \mathrm{Mpc}$. Everything above the dashed curves is excluded with $95 \, \% $ confidence. 
	\label{fig:ScreeningWithoutPrior}}
\end{figure*}

\begin{figure*}[th]
    \centering
	\includegraphics[trim={0.1cm 0 0.7cm 0},width=0.34\linewidth,clip]{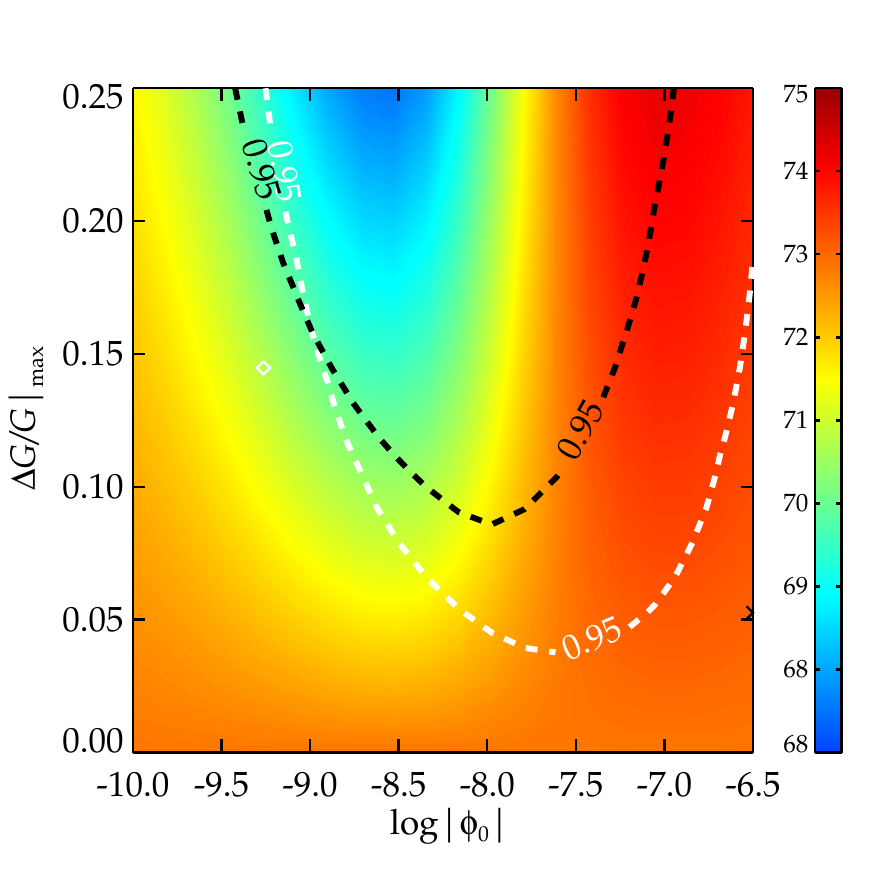}
    \includegraphics[trim={1.0cm 0 0.7cm 0},width=0.32\linewidth,clip]{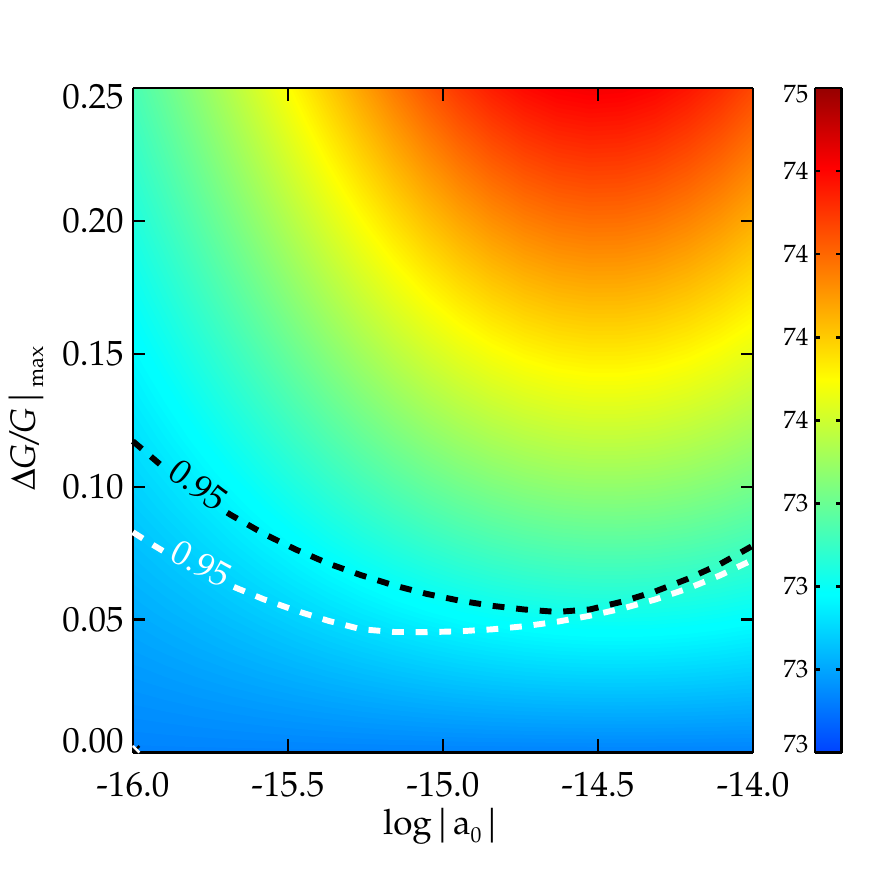}
    \includegraphics[trim={1cm 0 0.7cm 0},width=0.32\linewidth,clip]{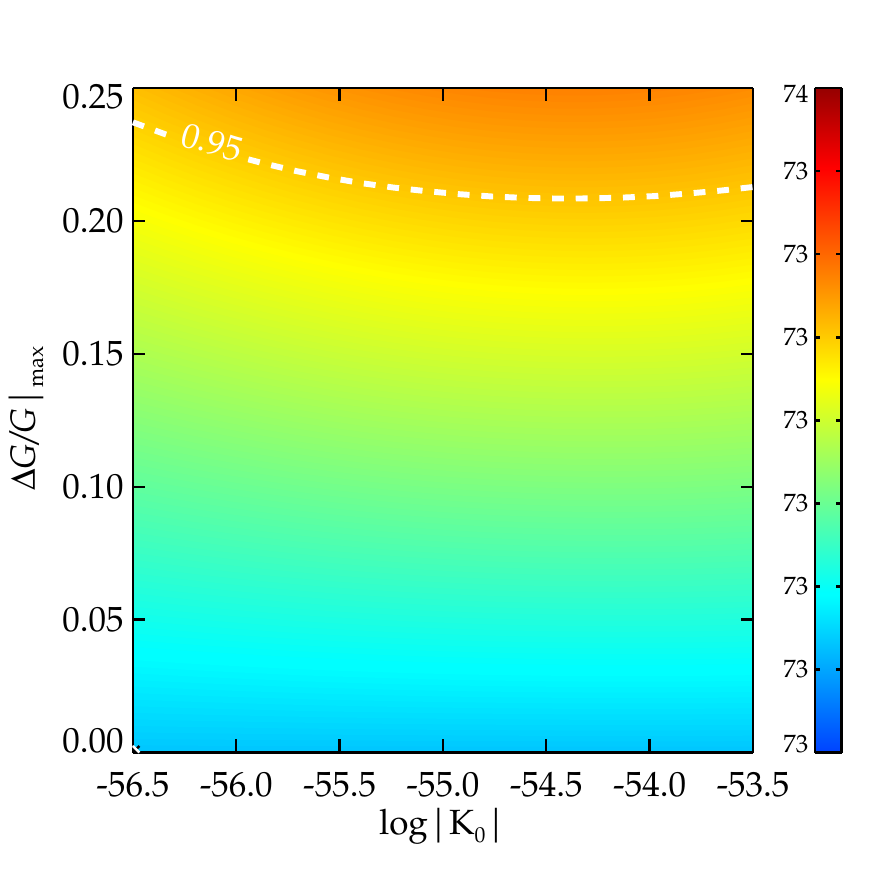}
    
	\caption{Results corresponding to Fig.~\ref{fig:ScreeningWithoutPrior}, but with a lower value of the transition parameter, $q=20$. (No weighting with the \emph{Planck} $H_0$). The confidence contours are widened and shifted upwards as discussed in the text.
	\label{fig:Lowq}}
\end{figure*}

\section{Calibrating the distance ladder without \emph{Planck}}
\label{sec:NoH0Prior}

\begin{table}
  \centering
  \begin{tabular}{c|c|c|c|c}
    \hline \hline
    & GR & $\Phi (0.4 \, \mathrm{Mpc})$ & $a (18.1 \, \mathrm{Mpc})$ & $K (5.1 \, \mathrm{Mpc})$\\
    \hline
    
    $H_0$ (best fit) & $73.2\pm 1.3$ & $72.3\pm 1.3$ & $72.7\pm 1.3$ & $72.9\pm 1.3$\\

    Tension & $4.1\sigma$ & $3.5\sigma$ & $3.8\sigma$ & $4.0\sigma$\\

    \hline

    $H_0$ (low) & -- & $70.9\pm 1.3$ & $71.9\pm 1.3$ & $72.2\pm 1.3$\\

    Tension & -- & $2.5\sigma$ & $3.2\sigma$ & $3.4\sigma$\\

\hline \hline
  \end{tabular}
  \caption{Effect of some example fifth force models on the Hubble tension (no weighting with the \emph{Planck} $H_0$). GR denotes the standard case without a fifth force. Here, $H_0$ (best fit) denotes the value of the Hubble constant at the best fit point in parameter space while $H_0$ (low) is the minimum value allowed by the $95 \, \%$ CLs from the distance ladder and the Cepheid-TRGB consistency. Here, the error bars on $H_0$ are set by the uncertainty in the determination of $H_0$ at the corresponding point in the parameter space.}
  \label{tab:resultsNoH0prior}
\end{table}

In this appendix we analyze the calibration of the cosmic distance ladder without marginalizing over proxy parameters. In this case, the probability of the fit to data at each point in the proxy parameter space does not include the tension to the \emph{Planck} value for $H_0$.
In Fig.~\ref{fig:ScreeningWithoutPrior}, we show the results corresponding to Fig.~\ref{fig:ScreeningWithPrior}.
As can be seen, without taking the tension to the \emph{Planck} $H_0$ into account, the excluded region in the parameter space assumes a ``U-shape'', so the allowed values of the model parameters $p_0$ and $\relGmax$ are unbounded; a large $\relGmax$ is admitted by letting $p_0$ be small enough (or large enough) and any value of $p_0$ is allowed as long as $\relGmax$ is small enough.
With the allowed region being unbounded in the parameter space, it is not possible to constrain the value of the Hubble constant 
for each proxy model, but rather only for each point in the proxy model parameter space.\footnote{Possibly, a one-sided constraint on $H_0$ could be obtained, although not in a Bayesian sense.} 

In Tab.~\ref{tab:resultsNoH0prior}, we show the results corresponding to Tab.~\ref{tab:resultsH0prior} for some example models. We see that the best-fit models only exhibit a minor improvement with respect to the $H_0$ tension and the minimum values for $H_0$ allowed by the $95 \, \%$ CLs (from the distance ladder and Cepheid-TRGB consistency) are in $>2 \sigma$ tension with \emph{Planck}. For some other models, it is possible to ease the tension to $\simeq 2 \sigma$ while satisfying the TRGB consistency test.
Yet, it is important to keep in mind that a definitive constraint on $H_0$ cannot be obtained for these models. The values in Tab.~\ref{tab:resultsNoH0prior} represent the values at the best-fit points (and the lowest values allowed) with the error bars on the Hubble constant reflecting the uncertainty in the determination of $H_0$ at this point in the parameter space.

\begin{figure*}[t]
    \centering
	\includegraphics[trim={0.1cm 0 0.7cm 0},width=0.34\linewidth,clip]{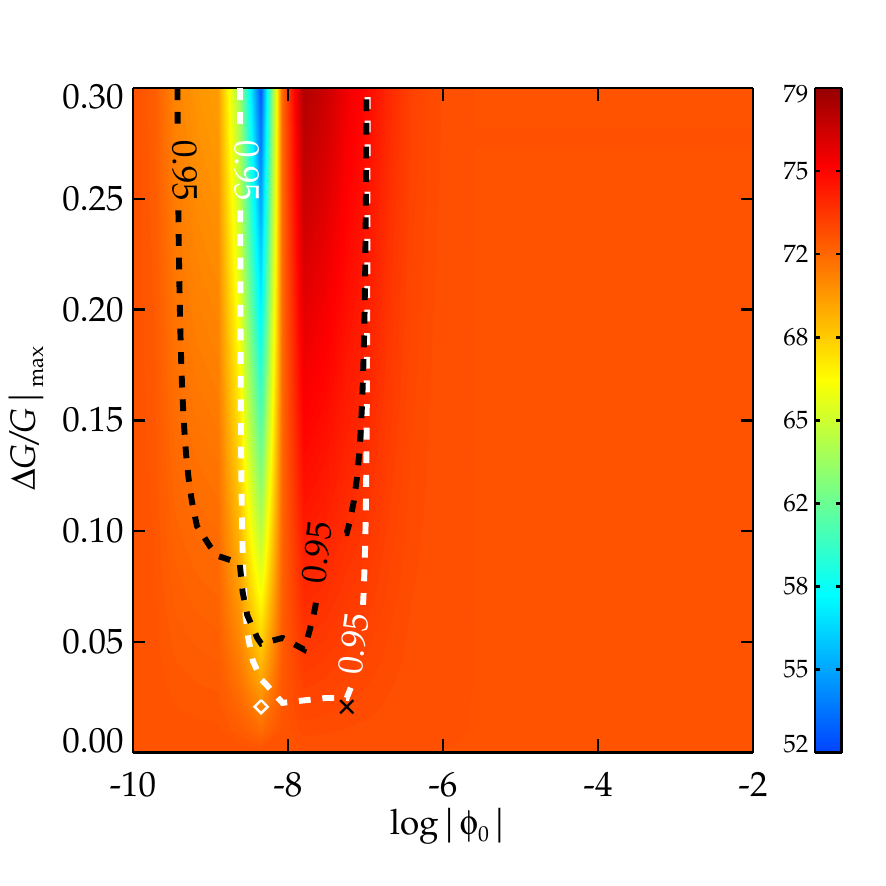}
    \includegraphics[trim={1.0cm 0 0.7cm 0},width=0.32\linewidth,clip]{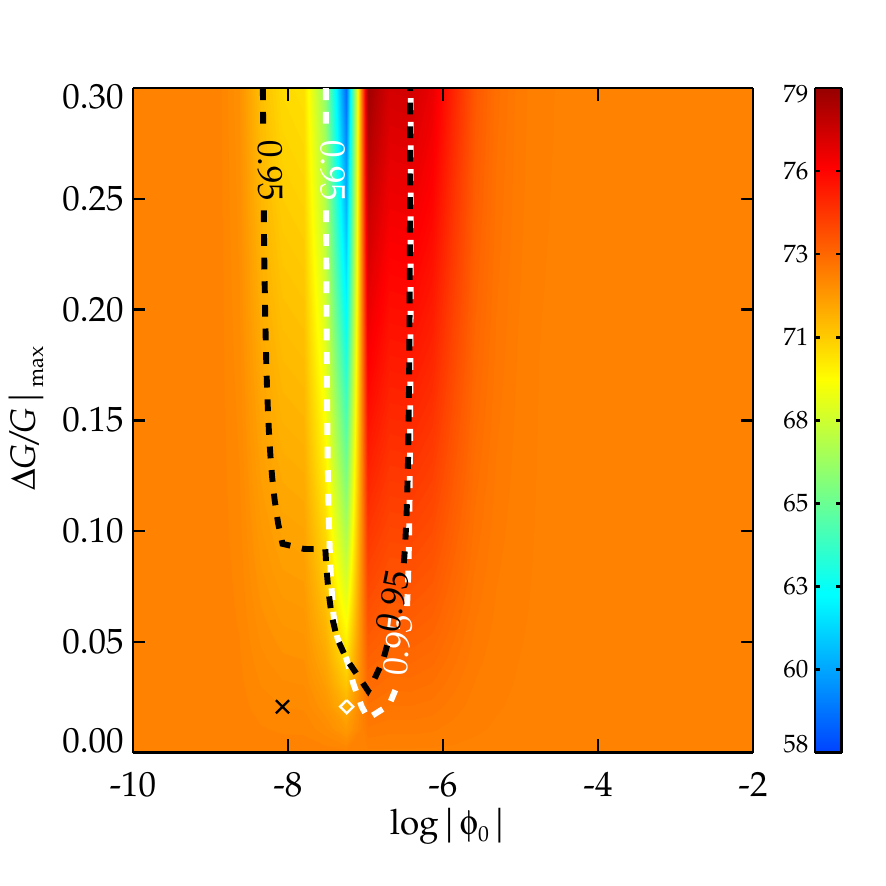}
    \includegraphics[trim={1cm 0 0.7cm 0},width=0.32\linewidth,clip]{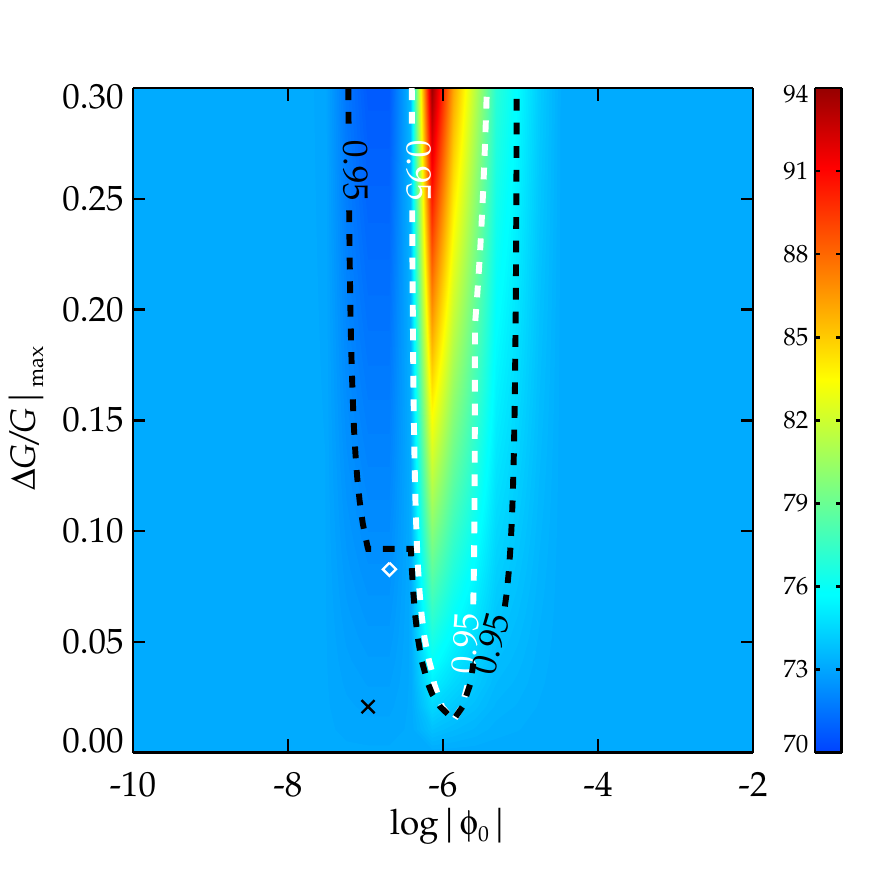}
    
	\caption{Inferred $H_0$ as a function of the model parameters for $\Phi$-screening models. \emph{Left}: $\Rmax = 0.4 \, \mathrm{Mpc}$. \emph{Middle}: $\Rmax = 1.4 \, \mathrm{Mpc}$. \emph{Right}: $\Rmax = 5.1 \, \mathrm{Mpc}$. Everything above the dashed curves is excluded with $95 \, \% $ confidence (no weighting with the \emph{Planck} $H_0$). The excluded region moves to the right as $\Rmax$ increases. 
	\label{fig:ScreeningWithoutPriorRmaxGrid}}
\end{figure*}

Changing the value of $q$ shifts the constraints in the parameter space in the vertical direction in the sense that smaller $q$ allows for a larger range of $\relGmax$. This is due to the fact that a small $q$ leads to a slow variation of $\relG$ with respect to the proxy value $p$. Hence, all galaxies obtain approximately the same value for $G$, resulting in no modification to the local distance ladder. At the same time, the widened transition between screened/unscreened galaxies also widens the excluded region in the $(p_0 ,\relGmax)$-plane. To summarize, a decrease to $q=20$ widens the ``U-shape'' of Fig.~\ref{fig:Lowq} and pushes it upwards compared with the default case in Fig.~\ref{fig:ScreeningWithoutPrior} where $q=500$.

With increasing $\Rmax$, we increase the range out to which the fifth force from a source contributes, thereby increasing the proxy value $p$ for each galaxy. Accordingly, we expect the excluded region in the parameter space to be pushed towards larger values of $p_0$.  In other words, the excluded ``U-shape'' is pushed to the right. The shape of the excluded region also changes to some degree due to the fact that different galaxies change their proxy values by different amounts when $\Rmax$ varies, see Fig.~\ref{fig:ScreeningWithoutPriorRmaxGrid}.

\begin{table}[h]
  \centering
  \begin{tabular}{ccc|ccc}
    \hline \hline
    Proxy & $\Rmax$ & Correlation & $H_0$ & $p$-value & TRGB $\chi^2$\\
    \hline
    
    \multirow{2}{*}{$\Phi$} & \multirow{2}{*}{$0.4 \, \mathrm{Mpc}$} & corr. & $68.0\pm 1.3$ & $0.44$ & $33$\\

    &  & uncorr. & $68.1\pm 1.3$ & $0.41$ & $34$\\

    \multirow{2}{*}{$a$} & \multirow{2}{*}{$0.4 \, \mathrm{Mpc}$} & corr. & $70.2\pm 1.4$ & $0.19$ & $20$\\

    &  & uncorr. & $70.2\pm 1.5$ & $0.21$ & $20$\\

    \multirow{2}{*}{$K$} & \multirow{2}{*}{$0.4 \, \mathrm{Mpc}$} & corr. & $71.1\pm 1.4$ & $0.09$ & $21$\\

    &  & uncorr. & $70.9\pm 1.7$ & $0.09$ & $21$\\

    \hline

    \multirow{2}{*}{$\Phi$} & \multirow{2}{*}{$1.4 \, \mathrm{Mpc}$} & corr. & $68.6\pm 1.3$ & $0.34$ & $37$\\

    &  & uncorr. & $68.7\pm 1.4$ & $0.31$ & $36$\\

    \multirow{2}{*}{$a$} & \multirow{2}{*}{$1.4 \, \mathrm{Mpc}$} & corr. & $70.6\pm 1.3$ & $0.14$ & $19$\\

    &  & uncorr. & $70.9\pm 1.6$ & $0.15$ & $20$\\

    \multirow{2}{*}{$K$} & \multirow{2}{*}{$1.4 \, \mathrm{Mpc}$} & corr. & $69.9\pm 1.3$ & $0.18$ & $19$\\

    &  & uncorr. & $70.1\pm 1.6$ & $0.17$ & $20$\\

    \hline

    \multirow{2}{*}{$\Phi$} & \multirow{2}{*}{$5.1 \, \mathrm{Mpc}$} & corr. & $71.3\pm 1.4$ & $0.07$ & $21$\\

    & & uncorr. & $71.7\pm 1.4$ & $0.08$ & $21$\\

    \multirow{2}{*}{$a$} & \multirow{2}{*}{$5.1 \, \mathrm{Mpc}$} & corr. & $70.5 \pm 1.3$ & $0.13$ & $21$\\

    &  & uncorr. & $70.8\pm 1.6$ & $0.14$ & $20$\\

    \multirow{2}{*}{$K$} & \multirow{2}{*}{$5.1 \, \mathrm{Mpc}$} & corr. & $71.3\pm 1.3$ & $0.07$ & $17$\\

    &  & uncorr. & $71.2\pm 1.5$ & $0.08$ & $18$\\

    \hline

    \multirow{2}{*}{$\Phi$} & \multirow{2}{*}{$18.1 \, \mathrm{Mpc}$} & corr. & $70.1\pm 1.3$ & $0.13$ & $21$\\

    & & uncorr. & $70.8\pm 1.6$ & $0.13$ & $23$\\

    \multirow{2}{*}{$a$} & \multirow{2}{*}{$18.1 \, \mathrm{Mpc}$} & corr. & $70.2\pm 1.3$ & $0.15$ & $21$\\

    &  & uncorr. & $70.3\pm 1.4$ & $0.14$ & $21$\\

    \multirow{2}{*}{$K$} & \multirow{2}{*}{$18.1 \, \mathrm{Mpc}$} & corr. & $70.9\pm 1.3$ & $0.09$ & $22$\\

    &  & uncorr. & $71.0\pm 1.4$ & $0.09$ & $22$\\

    \hline

    \multirow{2}{*}{$\Phi$} & \multirow{2}{*}{$50 \, \mathrm{Mpc}$} & corr. & $70.4\pm 1.3$ & $0.11$ & $21$\\

    & & uncorr. & $71.6\pm 2.0$ & $0.12$ & $42$\\

    \multirow{2}{*}{$a$} & \multirow{2}{*}{$50 \, \mathrm{Mpc}$} & corr. & $69.4\pm 1.3$ & $0.25$ & $34$\\

    &  & uncorr. & $69.3\pm 1.4$ & $0.25$ & $38$\\

    \multirow{2}{*}{$K$} & \multirow{2}{*}{$50 \, \mathrm{Mpc}$} & corr. & $71.1\pm 1.3$ & $0.08$ & $21$\\

    &  & uncorr. & $71.2\pm 1.4$ & $0.08$ & $22$\\

\hline \hline
  \end{tabular}
  \caption{Comparing results for the \emph{Planck}-weighted local distance ladder with completely correlated versus completely uncorrelated errors in the proxy values.}
  \label{tab:CorrUncorr}
\end{table}

\section{Uncorrelated proxy errors}
\label{sec:UncorrErr}
In the main text we have assumed that the errors in the proxy values are completely correlated. Another assumption would be that the errors in the proxy values are completely uncorrelated. Most likely, the truth lies somewhere in between these two extremes. However, a comparison shows that the choice of correlation does not significantly affect our conclusions concerning the Hubble tension. Typically, the difference in $H_0$ between correlated and uncorrelated errors is a few tenths of a $\mathrm{km/s/Mpc}$ and the differences in $p$ is typically $\simeq 0.01$. See Tab.~\ref{tab:CorrUncorr} for a comparative list.

\section{Complementary results}
\label{sec:ComplRes}
In Tab.~\ref{tab:resultsComp}, we present a comprehensive list of numerical results for the screening models, including the \emph{Planck} $H_0$ as a weight factor in the local distance ladder when marginalizing over model parameters.

\newpage

\begin{table*}[thb]
\renewcommand*{\arraystretch}{1.6}
  \centering
  \begin{tabular}{c|c|c|c|c|c|c|c|c|c|c}
    \hline\hline
    Proxy & $\Rmax$ & $\left. \frac{\Delta G}{\GN} \right|_\mathrm{max}$ & $\log_{10} |p_0|$ & $\left\langle \frac{\Delta G}{\GN} \right\rangle$ & Unscr.frac. & Anch.scr. & $H_0$ & $p$-value & $\chi^2$ & TRGB $\chi^2$\\[0.7ex]
    \hline
    GR & -- & -- & -- & -- & -- & -- & $73.2 \pm 1.3^*$ & $0.01$ & $1747$ & $20.8$\\
    \hline
    \multirow{5}{*}{$\Phi$} & $0.4 \, \mathrm{Mpc}$ & $0.05$ & $-8.3$ & $0.02$ & $58 \, \%$ & $1/1$ & $68.0 \pm 1.3$  & $0.44$ & $1622$ & $32.8$\\
    & $1.4 \, \mathrm{Mpc}$ & $0.06$ & $-7.2$ & $0.03$ & $47 \, \%$ & $1/1$ & $68.6 \pm 1.3$ & $0.34$ & $1638$ & $36.6$\\
    & $ 5.1 \, \mathrm{Mpc}$ & $0.16$ & $-7.1$ & $0.01$ & $5 \, \%$ & $1/1$ & $71.3 \pm 1.4$ & $0.07$ & $1699$ & $20.8$\\
    & $ 18.1 \, \mathrm{Mpc}$ & $0.09$ & $-5.2$ & $0.02$ & $21 \, \%$ & $1/1$ & $ 70.1\pm 1.3$ & $0.13$ & $1679$ & $21.3$\\
    & $50 \, \mathrm{Mpc}$ & $0.09$ & $-4.2$ & $0.01$ & $16 \, \%$ & $1/1$ & $70.4 \pm 1.3$ & $0.11$ & $1686$ & $20.6$\\
    \hline
    \multirow{5}{*}{$a$} & $0.4 \, \mathrm{Mpc}$ & $0.13$ & $-15.6$ & $0.02$ & $21 \, \%$ & $1/1$ & $ 70.2\pm1.4 $ & $0.19$ & $1664$ & $20.3$\\
    & $ 1.4\, \mathrm{Mpc}$ & $0.14$ & $-15.7$ & $0.01$ & $16 \, \%$ & $1/1$ & $ 70.6\pm1.3 $ & $0.14$ & $1677$ & $19.0$\\
    & $5.1 \, \mathrm{Mpc}$ & $0.13$ & $-15.6$ & $0.02$ & $16 \, \%$ & $1/1$ & $70.5 \pm 1.3$ & $0.13$ & $1678$ & $20.8$\\
    & $18.1 \, \mathrm{Mpc}$ & $0.09$ & $-15.2$ & $0.02$ & $26 \, \%$ & $1/1$ & $70.2 \pm 1.3 $ & $0.15$ & $1674$ & $20.9$\\
    & $50 \, \mathrm{Mpc}$ & $0.09$ & $-14.7$ & $0.02$ & $26 \, \%$ & $1/1$ & $69.4 \pm 1.3$ & $0.25$ & $1654$ & $33.7$\\
    \hline
    \multirow{5}{*}{$K$} & $0.4 \, \mathrm{Mpc}$ & $0.17$ & $-55.5$ & $0.02$ & $42 \, \%$ & $1/1$ & $71.1 \pm 1.4$ & $0.09$ & $1690$ & $20.9$\\
    & $ 1.4\, \mathrm{Mpc}$ & $0.13$ & $-55.4$ & $0.02$ & $37 \, \%$ & $1/1$ & $ 69.9\pm1.3 $ & $0.18$ & $1667$ & $19.5$\\
    & $5.1 \, \mathrm{Mpc}$ & $0.17$ & $-55.4$ & $0.01$ & $26 \, \%$ & $1/1$ & $ 71.3\pm1.3 $ & $0.07$ & $1698$ & $16.6$\\
    & $18.1 \, \mathrm{Mpc}$ & $0.16$ & $-55.4$ & $0.02$ & $26 \, \%$ & $1/1$ & $ 70.9\pm1.3 $ & $0.09$ & $1692$ & $21.6$\\
    & $50 \, \mathrm{Mpc}$ & $0.17$ & $-55.3$ & $0.02$ & $21 \, \%$ & $1/1$ & $71.1 \pm1.3 $ & $0.08$ & $1697$ & $21.2$\\
    \hline\hline
  \end{tabular}
  \caption{The effect of various screening models on the Hubble tension (with the \emph{Planck}-weighted local distance ladder) and the TRGB consistency test. Here, $q=500$. The tabulated $\relGmax$ and $p_0$ are the best-fit values. The unscreening fraction denotes the fraction of host galaxies with $\relG > 0.01$. The anchor screening is entered on the format $n/m$ where the first slot indicates MW and LMC and the second slot N4258. Here, ``0'' stands for no screening (i.e., $\relG > 0.01$) and ``1'' stands for screened. The unscreening fraction and anchor screening are calculated at the best-fit point. The $p$-values and $\chi^2$-values are obtained from the quality of the global distance ladder fit with the \emph{Planck} $H_0$ added. This table should be considered as a guide to the qualitative behavior of the fifth force models rather than a list of exact numbers. To save computational time, we have used a coarser grid in the parameter space here than in the results featured in the main text. $^*$For reference, here we have entered the standard $H_0$ value in the case of no fifth force, obtained from the distance ladder alone.}
  \label{tab:resultsComp}
\end{table*}

\bibliography{bibliography}{}
\bibliographystyle{apsrev4-1}

\end{document}